\pdfoutput=1

\documentclass[11pt,twoside,a4paper,cmspaper,final,collab]{cms-tdr}
\def\svnVersion{231014}\def\cmsCernNo{2013-037}\def\cmsCernDate{\today}\def\cmsMessage{Published in the Journal of High Energy Physics as \href{http://dx.doi.org/10.1007/JHEP01(2013)077}{\doi{10.1007/JHEP01(2013)077.}}}

\begin{document}\cmsNoteHeader{SUS-11-022}

\hyphenation{had-ron-i-za-tion}
\hyphenation{cal-or-i-me-ter}
\hyphenation{de-vices}
\RCS$Revision: 160355 $
\RCS$HeadURL: svn+ssh://alverson@svn.cern.ch/reps/tdr2/papers/SUS-11-022/trunk/SUS-11-022.tex $
\RCS$Id: SUS-11-022.tex 160355 2012-12-06 13:19:33Z bainbrid $

\newcommand{\rat}{\ensuremath{R_{\alpha_{\mathrm{T}}}}\xspace}
\newcommand{\RaT}{\rat}
\newcommand{\sq}{\ensuremath{\tilde{\mathrm{q}}}\xspace}
\newcommand{\st}{\ensuremath{\tilde{\mathrm{t}}}\xspace}
\newcommand{\gl}{\ensuremath{\tilde{\mathrm{g}}}\xspace}

\newcommand{\scalht}{\mbox{$H_\mathrm{T}$}\xspace}
\newcommand{\eslash}{{\hbox{$E$\kern-0.6em\lower-.05ex\hbox{/}\kern0.10em}}}
\newcommand{\met}{\mbox{$\eslash_\mathrm{T}$}\xspace}
\newcommand{\cls}{\mbox{CL$_s$}\xspace}
\newcommand{\znunu}{\ensuremath{\cPZ \rightarrow \cPgn\cPagn}}
\newcommand{\wtaunu}{\ensuremath{\PW \rightarrow \Pgt\cPgn}}
\newcommand{\Et}{\ensuremath{{E_{\text T}}}\xspace}
\newcommand{\Hslash}{{\hbox{$H$\kern-0.8em\lower-.05ex\hbox{/}\kern0.10em}}}
\newcommand{\mht}{\mbox{$\Hslash_\mathrm{T}$}\xspace}
\newcommand{\dht}{\ensuremath{\Delta\scalht}\xspace}

\newcommand{\alphat}{\ensuremath{\alpha_{\text{T}}}\xspace}
\newcommand{\htalphat}{\texttt{HT\_AlphaT}\xspace}
\newcommand{\photon}{\texttt{Photon}\xspace}
\newcommand{\muht}{\texttt{Mu\_HT}\xspace}
\newcommand{\httrigger}{\texttt{HT}\xspace}
\newcommand{\htmht}{\texttt{HT\_MHT}\xspace}
\newcommand{\mt}{\ensuremath{M_\mathrm{T}}\xspace}
\newcommand{\gj}{\ensuremath{\gamma} + jets\xspace}
\newcommand{\mj}{\ensuremath{\mu} + jets\xspace}
\newcommand{\mmj}{\ensuremath{\mu\mu} + jets\xspace}
\newcommand{\npre}{\ensuremath{N_{\textrm{pred}}}\xspace}
\newcommand{\nobs}{\ensuremath{N_{\textrm{obs}}}\xspace}
\newcommand{\njets}{\ensuremath{N_{\textrm{jet}}}\xspace}

\newlength\cmsFigWidth
\ifthenelse{\boolean{cms@external}}{\setlength\cmsFigWidth{0.85\columnwidth}}{\setlength\cmsFigWidth{0.4\textwidth}}
\ifthenelse{\boolean{cms@external}}{\providecommand{\cmsLeft}{top}}{\providecommand{\cmsLeft}{left}}
\ifthenelse{\boolean{cms@external}}{\providecommand{\cmsRight}{bottom}}{\providecommand{\cmsRight}{right}}

\cmsNoteHeader{SUS-11-022}

\title{Search for supersymmetry in final states with missing
transverse energy and 0, 1, 2, or ${\ge}3$ b-quark jets in 7\TeV pp
collisions using the variable \alphat}

\date{\today}

\abstract{A search for supersymmetry in final states with jets and
missing transverse energy is performed in pp collisions at a
centre-of-mass energy of $\sqrt{s} = 7$\TeV. The data sample
corresponds to an integrated luminosity of 4.98\fbinv collected by
the CMS experiment at the LHC. In this search, a dimensionless
kinematic variable, $\alpha_\mathrm{T}$, is used as the main
discriminator between events with genuine and misreconstructed missing
transverse energy. The search is performed in a signal region that is
binned in the scalar sum of the transverse energy of jets and the
number of jets identified as originating from a bottom quark. No
excess of events over the standard model expectation is found.
Exclusion limits are set in the parameter space of the constrained
minimal supersymmetric extension of the standard model, and also in
simplified models, with a special emphasis on compressed spectra and
third-generation scenarios.}

\hypersetup{
pdfauthor={CMS Collaboration},
pdftitle={Search for supersymmetry in final states with missing
transverse energy and 0, 1, 2, or at least 3 b-quark jets in 7 TeV pp
collisions using the variable alphaT},
pdfsubject={CMS, physics, jets, missing energy, supersymmetry, SUSY, AlphaT},
pdfkeywords={CMS, physics, jets, missing energy, supersymmetry, SUSY, AlphaT},
}

\maketitle

\section{Introduction}

Supersymmetry (SUSY) is generally regarded as one of the likely
extensions to the standard model (SM) of particle
physics~\cite{ref:SUSY-1, ref:SUSY0, ref:SUSY1, ref:SUSY2, ref:SUSY3,
  ref:SUSY4, ref:hierarchy1, ref:hierarchy2}. It is based on the
unique extension of the space-time symmetry group underpinning the SM,
introducing a relationship between fermions and bosons. A low-energy
realisation of SUSY, e.g. at the TeV scale, is motivated by the
cancellation of the quadratically divergent loop corrections to the
Higgs boson mass in the SM~\cite{ref:hierarchy1,
  ref:hierarchy2}. These corrections are proportional to the masses of
the particles that couple to the Higgs boson. The most relevant terms
come from the interplay between the masses of the third generation
(top and bottom) squarks, and the largest Yukawa coupling (of the top
quark).

In order to avoid large cancellations in these loop corrections, the
difference in masses between the top quark and the third generation
squarks must not be too large~\cite{ref:barbierinsusy}. While the
majority of SUSY particles might not be accessible at the present
energy and luminosity delivered by the Large Hadron Collider (LHC),
the recent discovery of a low-mass Higgs boson
candidate~\cite{ref:atlashiggsdiscovery, ref:cmshiggsdiscovery}
motivates models in which top and bottom squarks appear at the
TeV scale. Furthermore, if the multiplicative quantum number
R-parity~\cite{Farrar:1978xj} is conserved, SUSY particles will be
produced in pairs and decay to SM particles and the lightest SUSY
particle (LSP), which is generally assumed to be weakly interacting
and massive. This would result in a final state that is rich in jets,
especially those originating from bottom quarks, and contains a
significant amount of missing transverse energy, \met.

This paper summarises a search that is designed to be sensitive to
missing transverse energy signatures in events with two or more
energetic jets that are categorised according to the number of
reconstructed jets originating from bottom quarks (b-quark jets) per
event. With respect to previous searches~\cite{RA1Paper2011,RA1Paper},
this refinement provides improved sensitivity to third generation
squark signatures. However, the same inclusive search strategy is
deployed, thus maintaining the ability to identify a wide variety of
SUSY event topologies arising from the pair production and decay of
massive coloured sparticles.

The ATLAS and Compact Muon Solenoid (CMS) experiments have performed
various searches \cite{RA1Paper2011, RA1Paper, cms1, cms2, cms3,
  atlas2, atlas3, atlas4, atlas1} for the production of massive
coloured sparticles and their subsequent decay to a final state of
jets and missing transverse energy. These searches were performed with
a dataset of pp collisions at $\sqrt{s} = 7$\TeV, and no significant
deviations from SM expectations were observed. The majority of these
searches have been interpreted in the context of a specific model of
SUSY breaking, the constrained minimal supersymmetric extension of the
standard model (CMSSM)~\cite{ref:MSUGRA, PhysRevLett.69.725,
  ref:CMSSM}.
The simplifying assumption of this model is universality at an energy scale of
${\cal{O}}(10^{16})\GeV$ which makes the CMSSM a useful framework to study
SUSY phenomenology at colliders, and to serve as a benchmark for the
performance of experimental searches.

However, the universality conditions of the CMSSM result in
significant restrictions on the possible SUSY particle mass spectra
and thus kinematic signatures. This limits the interpretation of the
results in scenarios such as the direct production of third-generation
squarks and compressed spectra, where the mass difference between the
primary produced sparticle (\eg, a squark or a gluino) and the LSP is
small. Therefore, in order to complement the interpretation within the
CMSSM, simplified models~\cite{Alwall:2008ag,Alwall:2008va,sms} are
also used to interpret the search results. These models are
characterised using a limited set of SUSY particles (production and
decay) and enable comprehensive studies of individual SUSY event
topologies. The simplified model studies can be performed without
limitations on fundamental properties such as decay modes,
production cross sections, and sparticle masses. A special emphasis is
placed on interpretation within models involving compressed spectra or
third generation squarks.

\section{The CMS apparatus\label{sec:cms}}

The central feature of the CMS detector is a superconducting solenoid,
which provides an axial magnetic field of 3.8\unit{T}. The bore of the
solenoid is instrumented with several particle detection
systems. Silicon pixel and strip tracking systems measure charged
particle trajectories with full azimuthal ($\phi$) coverage and a
pseudorapidity acceptance of $|\eta| < 2.5$, where $\eta = -\ln [ \tan
(\theta/2) ]$ and $\theta$ is the polar angle with respect to the
counterclockwise beam direction. The resolutions on the transverse
momentum (\pt) and impact parameter of a charged particle with $\pt <
40\gev$ are typically 1\% and 15\mum, respectively. A
lead-tungstate crystal electromagnetic calorimeter (ECAL) and a
brass/scintillator hadron calorimeter surround the tracking
volume. The region outside the solenoid is covered by an
iron/quartz-fiber hadron calorimeter. The ECAL covers $|\eta| < 3.0$
and provides an energy resolution of better than 0.5\% for unconverted
photons with transverse energies above 100\gev. The hadron
calorimeters cover $|\eta| < 5.0$ with a resolution in jet energy, E
(GeV), of about 100\%/$\sqrt{\textrm{E}}$ for the region $|\eta| <
3.0$. Muons are identified in gas-ionization detectors, covering
$|\eta| < 2.4$, embedded in the steel return yoke. The CMS detector is
nearly hermetic, which allows momentum-balance measurements in the
plane transverse to the beam axis. A two-tier trigger system is
designed to select the most interesting pp collision events for use in
physics analysis. A detailed description of the CMS detector can be
found elsewhere~\cite{ref:CMS}.

\section{Object definitions and event reconstruction\label{sec:event}}

The event reconstruction and selection criteria follow the procedure
described in Refs.~\cite{RA1Paper,RA1Paper2011}.  Jets are
reconstructed from energy deposits in the calorimeters, clustered by
the anti-\kt algorithm~\cite{antikt} with a distance parameter
of 0.5. The raw jet energies measured by the calorimeter systems are
corrected to establish a uniform relative response in $\eta$ and a
calibrated absolute response in transverse momentum with an associated
uncertainty between 2\% and 4\%, depending on the jet $\eta$ and
\pt~\cite{Chatrchyan:2011ds}. Jets considered in the analysis are
required to have transverse energy $\Et > 50\GeV$ and the two
highest-$\Et$ jets must each satisfy $\Et > 100\GeV$. These two \Et\
requirements change under special circumstances described in
Section~\ref{sec:alphat}. The highest-$\Et$ jet is additionally
required to be within the central tracker acceptance ($|\eta| < 2.5$).
Events are vetoed if any additional jet satisfies both $\Et > 50\GeV$
and $|\eta| > 3$, or rare, spurious signals are identified in the
calorimeters~\cite{1748-0221-5-03-T03014}. To
suppress SM processes with genuine \met from neutrinos in the final
state, an event is vetoed if it contains an isolated
electron~\cite{PAS-EGM-10-004} or muon~\cite{PAS-MUO-10-004} with $\pt
> 10\GeV$. Further, events with an isolated photon~\cite{PAS-EGM-10-006} with $\pt >
25\GeV$ are also vetoed.

The presence of a b-quark jet is identified through a vertex that is
displaced with respect to the primary interaction, using an algorithm
that attempts to reconstruct a secondary vertex using tracks from
charged particles associated to each jet. Using a likelihood ratio
technique, the combined secondary vertex
algorithm~\cite{CMS-PAS-BTV-11-004} incorporates several variables
related to the vertex, such as decay length significance, mass, and
track multiplicity, to build a discriminator that distinguishes between jets originating
from bottom quarks and those from other sources. These include jets from charm
quarks (c-quark jets) and light-flavour quarks.  The algorithm also
provides a value for this discriminator based on single-track
properties, when no secondary vertices have been reconstructed.
Discriminator values above a certain threshold are used to tag jets as
originating from b quarks. This threshold is chosen such that the
mistagging rate, the probability to tag a jet originating from a
light-flavour quark, is approximately 1\% for jets with transverse
momenta of 80~GeV~\cite{CMS-PAS-BTV-11-003,CMS-PAS-BTV-11-004}. The
same threshold results in a b-tagging efficiency, the probability
to correctly tag a jet as originating from a bottom quark, in the
range 60--70\%~\cite{CMS-PAS-BTV-11-003,CMS-PAS-BTV-11-004}.

The following two variables characterise the visible energy and
missing momentum in the transverse plane: the scalar sum of the
transverse energy $\Et$ of jets, defined as $\scalht =
\sum_{i=1}^{\njets} \Et^{\mathrm{j}_i}$, and the magnitude of the vector sum of
the transverse momenta $\vec{\pt}$ of jets, defined as $\mht =
|\sum_{i=1}^{\njets} \vec{\pt}^{\mathrm{j}_i}|$, where $\njets$ is the
number of jets above the \Et\ threshold. Significant hadronic activity
in the event is ensured by requiring $\scalht > 275\GeV$.  Following
these selections, the background from multijet production, a
manifestation of quantum chromodynamics (QCD), is still several orders
of magnitude larger than the typical yields expected from a SUSY
signal.

\section{Selecting events with missing transverse energy\label{sec:alphat}}

The \alphat kinematic variable~\cite{Randall:2008rw, RA1Paper2011} is used to efficiently reject
multijet events without significant \met, including those with transverse
energy mismeasurements, while retaining a large sensitivity to new
physics with genuine \met signatures. For dijet events, the \alphat
variable is defined as:

\begin{equation}
\label{eq:alphat}
\alphat\, =\, \frac{\Et^{\mathrm{j}_2}}{M_\text{T}} \quad , \quad M_\text{T}\,
= \,\sqrt{ \left( \sum_{i=1}^2 \Et^{\mathrm{j}_i} \right)^2 - \left(
\sum_{i=1}^2 p_x^{\mathrm{j}_i} \right)^2 - \left( \sum_{i=1}^2 p_y^{\mathrm{j}_i} \right)^2}.
\end{equation}

where $\Et^\mathrm{j_2}$ is the transverse energy of the less energetic
jet, and $M_\mathrm{T}$ is the transverse mass of the dijet system. For
a perfectly measured dijet event with $\Et^\mathrm{j_1} = \Et^\mathrm{j_2}$
and jets back-to-back in $\phi$, and in the limit in which each jet's
momentum is large compared with its mass, the value of \alphat is
0.5. In the case of an imbalance in the measured transverse energies
of back-to-back jets, \alphat is smaller than 0.5. Values
significantly greater than 0.5 are observed when the two jets are not
back-to-back, recoiling against genuine \met.

For events with three or more jets, an equivalent dijet system is
formed by combining the jets in the event into two pseudo-jets. The
$\Et$ of each of the two pseudo-jets is calculated as the scalar sum
of the measured $\Et$ of the contributing jets. The combination chosen
is the one that minimises the $\Et$ difference (\dht) between the two
pseudo-jets. This simple clustering criterion provides the best
separation between multijet events and events with genuine
\met. Thus, in the case of events with at least three jets, the
\alphat variable can be defined as:

\begin{equation}
\label{eq:alphat2}
\alphat\, = \,\frac{1}{2}\cdot\frac{\scalht - \dht}{\sqrt{\scalht^2 -
\mht^2}} \quad = \,\frac{1}{2}\cdot\frac{1 -
(\dht/\scalht)}{\sqrt{1 - (\mht/\scalht)^2}}
\end{equation}

Events with extremely rare but large stochastic fluctuations in the
calorimetric measurements of jet energies can lead to values of
$\alpha_{\rm T}$ slightly above 0.5. Such events are rejected by
requiring $\alphat > 0.55$. A similar behaviour is observed in events
with reconstruction failures, severe energy losses due to detector
inefficiencies, or jets below the \Et threshold that result in
significant \mht relative to the value of \met (as measured by the
calorimeter systems, which is not affected by jet \Et
thresholds). These classes of events are rejected by applying
dedicated vetoes, described further in Ref.~\cite{RA1Paper}. The
leakage above 0.5 becomes smaller with increasing \scalht due to the
increase in average jet energy and thus an improvement in jet energy
resolution. Further, the relative impact of jets falling below the \Et
threshold is reduced as the energy scale of the event (\ie \scalht)
increases.

The signal region is defined by $\scalht > 275\GeV$ and $\alphat >
0.55$, which is divided into eight bins in \scalht: two bins of width
$50\GeV$ in the range $275 < \scalht < 375\GeV$, five bins of width
$100\GeV$ in the range $375 < \scalht < 875\GeV$, and a final open
bin, $\scalht > 875\GeV$. As in Ref.~\cite{RA1Paper}, the jet $\Et$
threshold is scaled for the two lowest \scalht bins
leading to thresholds of $37\GeV$ and $43\GeV$. The two highest-\Et jet
thresholds are scaled to $73\GeV$ and $87\GeV$. 
This approach maintains SM background admixtures and event kinematics
similar to those observed for the higher \scalht bins. Events are
further categorised according to whether they contain exactly zero,
one, two, or at least three reconstructed b-quark jets.

Events in the signal region are recorded with a dedicated trigger
condition that must satisfy simultaneously the requirements $\scalht >
250\gev$ and $\alphat > 0.53$, with the latter threshold increasing to
$0.60$ towards the end of 2011 due to higher instantaneous
luminosities. The efficiency with which events that would satisfy the
signal region selection criteria also satisfy the trigger conditions is
measured in data to be $(82.8 \pm 1.1)\%$, $(95.9 \pm 0.9)\%$, and
$(>98.5 \pm 0.9)\%$ for the regions $275 < \scalht < 325\gev$, $325 <
\scalht < 375\gev$, and $\scalht > 375\gev$, respectively.

A disjoint hadronic control sample consisting predominantly of
multijet events is defined by inverting the \alphat requirement for a
given \scalht region, which is used primarily in the estimation of any
residual background from multijet events. These events are recorded
by a set of triggers with thresholds only in \scalht.

\section{Background estimation from data\label{sec:bkgd}}

Once all the signal region selection requirements have been imposed, the
contribution from multijet events is expected to be negligible.  The
remaining significant backgrounds in the signal region stem
from SM processes with genuine \met in the final state.  In the case
of events where no b-quark jets are identified, the largest backgrounds with
genuine \met arise from the production of W and Z bosons in
association with jets. The weak decay \znunu\ is the only significant
contribution from Z + jets events. For W + jets events, the two
dominant sources are leptonic W decays in which the lepton is not
reconstructed or fails the isolation or acceptance requirements, and
the weak decay $\wtaunu$ where the $\tau$ decays hadronically and is
identified as a jet. Contributions from SM processes such as
single-top, Drell-Yan, and diboson production are also expected. For
events with one or more reconstructed b-quark jets, \ttbar
production followed by semi-leptonic weak decays becomes the most
important single background source. For events with only one
reconstructed b-quark jet, the contribution of both W + jets and Z + jets
backgrounds are of a similar size to the \ttbar background.  For
events with two reconstructed b-quark jets, \ttbar production dominates,
while events with three or more reconstructed b-quark jets originate almost
exclusively from \ttbar events, in which at least one jet is
misidentified as originating from a bottom quark.

In order to estimate the contributions from each of these backgrounds,
three data control samples are used, which are binned in the same way
as the signal region. The irreducible background of \znunu\ + jets
events in the signal region is estimated from two independent
data samples of $\cPZ\rightarrow\mu\mu$ + jets and $\gamma$ + jets
events, both of which share the kinematic properties of\znunu\ + jets but have different
acceptances. The $\cPZ\rightarrow\mu\mu$ + jets events have identical
kinematic properties to the \znunu\ + jets background when the two
muons are ignored, but a smaller branching fraction, while the
$\gamma$ + jets events have similar kinematic properties when the
photon is ignored~\cite{RA1Paper2011,Bern:2011pa}, but a larger
production cross section. A $\mu$ + jets data sample provides an
estimate for all other SM backgrounds, which is dominated by \ttbar
and W production leading to W + jets final states.

The event selection criteria for the control samples are defined to
ensure that any potential contamination from multijet events is
negligible. Further, the same selection criteria also strongly
suppress contributions from a wide variety of SUSY models, including
those considered in this analysis. Any potential signal contamination
in the data control samples is accounted for in the fitting procedure
described in Section~\ref{sec:results}.

\subsection{Definition of data control samples}

The $\mu$ + jets sample is recorded using two different trigger
strategies, to account for evolving trigger conditions during the 2011
run. The hadronic trigger condition, combining \scalht and \alphat, is
used for the region $275 < \scalht < 375\GeV$. Here, the event
selection, following closely the prescription described in
Ref.~\cite{toppaper}, requires exactly one isolated muon that
satisfies stringent quality criteria, with $\pt > 10\GeV$ and $|\eta|
< 2.1$. In order for the trigger to be maximally efficient, the
requirement $\alphat > 0.55$ is also imposed. For the region $\scalht
> 375\GeV$, the trigger condition requires both a muon above a \pt
threshold as high as $40\GeV$ and $\scalht > 300\GeV$. The muon must
satisfy $\pt >45\GeV$ in order for the trigger to be maximally
efficient, at $(91.3 \pm 0.1)\%$. The requirement $\alphat > 0.55$ is
again imposed when zero b-quark jets are reconstructed per event. For events
in which at least one b-quark jet is reconstructed, no \alphat requirement
is used. This approach increases the statistical precision of
predictions derived from event samples containing b-quark jets, while the
impact of relaxing the \alphat requirement is tested with a dedicated
set of closure tests described in Section~\ref{sec:methodology}.

In addition to the requirements described above, further selection
criteria are applied. The transverse mass of the muon and \met system
must be larger than $30\GeV$ to ensure a sample rich in W bosons. The
muon is required to be separated from the closest jet in the event by
$\Delta \eta$ and $\Delta \phi$ such that the distance $\Delta R
\equiv \sqrt{(\Delta \eta)^2 + (\Delta \phi)^2} > 0.5$.  To ensure that
this sample is disjoint from the $\mu\mu$ + jets sample, the event is
rejected if a second muon candidate is identified that does not
satisfy all quality criteria or is non-isolated or is outside the
acceptance, and the two muon candidates have an invariant mass that is
within a window of $\pm 25\GeV$ around the mass of the Z boson.

The $\mu\mu$ + jets sample follows the same trigger strategy and muon
identification criteria as the $\mu$ + jets sample. The event
selection requires exactly two oppositely charged, isolated muons
satisfying stringent quality criteria, and an invariant mass within a
window of $\pm 25\GeV$ around the mass of the Z boson.  Each muon is
required to be separated from the nearest jet in the event by the
distance $\Delta R > 0.5$. The same \alphat requirements are used as
for the \mj sample.

The $\gamma$ + jets sample is selected using a dedicated photon
trigger condition requiring a localised energy deposit in the ECAL
with $\Et > 135\GeV$ that satisfies loose photon identification and
isolation criteria~\cite{PAS-EGM-10-006}. The event selection requires
$\scalht > 375\GeV$, $\alphat > 0.55$, and a single photon to be
reconstructed with $\Et > 150\GeV$, $|\eta| < 1.45$, satisfying tight
isolation criteria, and with a minimum distance to any jet of $\Delta
R > 1.0$. For these selection criteria, the photon trigger condition
is found to be fully efficient.

\subsection{Method and systematic studies\label{sec:methodology}}

The method used to estimate the SM background contributions in the
signal region relies on the use of transfer factors, which are
functions of \scalht and the number of b-quark jets per event, $n_\cPqb$, and
are computed separately for each data control sample.
These transfer factors are determined from simulation samples generated
with \MADGRAPH v4.22~\cite{madgraph} interfaced to \PYTHIA
6.4 tune Z2~\cite{pythia}, and the \GEANT4-based~\cite{GEANT4} CMS detector
simulation. Each factor is defined as the ratio of yields from
simulation in a given bin of the signal region, $N_{\rm
  MC}^{\rm signal}(\scalht,n_\cPqb)$ and the corresponding bin of
one control sample, $N_{\rm MC}^{\rm control}(\scalht,n_\cPqb)$. The
factors are used to translate the observed yield measured in a control
sample bin, $\nobs^{\rm control}(\scalht,n_\cPqb)$ into an
expectation for one or more SM background processes in the
corresponding bin of the signal region, $\npre^{\rm
  signal}(\scalht,n_\cPqb)$:

\begin{equation}
  \label{equ:pred-method}
  \npre^{\rm signal}(\scalht, n_\cPqb) = \nobs^{\rm
    control}(\scalht, n_\cPqb) \times \frac{N_{\rm
      MC}^{\rm signal}(\scalht, n_{\rm
      b})}{N_{\rm MC}^{\rm control}(\scalht, n_{\rm
      b})}.
\end{equation}

In order to maximise sensitivity to potential new physics signatures
in final states with multiple b-quark jets, a method that improves the
statistical power of the predictions from simulation, particularly for
$n_\cPqb \ge 2$, is employed. The distribution of $n_\cPqb$ is
estimated from generator-level information contained in the
simulation, namely the number of reconstruction-level jets matched to underlying b
quarks, $n_\cPqb^{\rm gen}$, and light quarks, $n_{\rm q}^{\rm
  gen}$, per event. All relevant combinations of $n_\cPqb^{\rm gen}$
and $n_{\rm q}^{\rm gen}$ are considered, and event counts are
recorded in bins of \scalht for each combination $N(n_\cPqb^{\rm
  gen},n_{\rm q}^{\rm gen}, \scalht)$.  The b-tagging efficiency, $\epsilon$,
and a flavour-averaged mistagging rate, $m$, are also determined from
simulation for each \scalht bin, with both quantities averaged over
jet $p_{\rm T}$ and $\eta$. Corrections are applied on a jet-by-jet
basis to both $\epsilon$ and $m$ in order to match the corresponding
measurements with data~\cite{CMS-PAS-BTV-11-003,
  CMS-PAS-BTV-11-004}. This information is sufficient to predict
$n_\cPqb$ and thus also determine the yield from simulation for a
given bin, $N(\scalht, n_\cPqb)$:

\begin{equation}
  \label{equ:btag-formula}
N(\scalht, n_\cPqb) = \sum_{n_{\textrm{b}}^{\textrm{gen}} + n_{\textrm{q}}^{\textrm{gen}} = N_{\textrm{jet}}} \; \sum_{n_{\textrm{b}}^{\textrm{tag}} + n_{\textrm{q}}^{\textrm{tag}} = n_{\textrm{b}}} N(n_{\textrm{b}}^{\textrm{gen}}, n_{\textrm{q}}^{\textrm{gen}}, \scalht) \times P(n_{\textrm{b}}^{\textrm{tag}} ; n_{\textrm{b}}^{\textrm{gen}}, \epsilon) \times P(n_{\textrm{q}}^{\textrm{tag}} ; n_{\textrm{q}}^{\textrm{gen}}, m)
\end{equation}

where $n_{\textrm{b}}^{\textrm{tag}}$ and $n_{\textrm{q}}^{\textrm{tag}}$ are the number of times a
reconstruction-level b-quark jet originates from an underlying b-quark and light-quark respectively,
and $P(n_{\textrm{b}}^{\textrm{tag}} ; n_{\textrm{b}}^{\textrm{gen}}, \epsilon)$ and $P(n_{\textrm{q}}^{\textrm{tag}} ; n_{\textrm{q}}^{\textrm{gen}}, m)$ are the binomial probabilities for this to happen.
The predicted yields are found to be in good agreement with
the yields obtained directly from the simulation in those bins with significant population.

The method exploits the ability to determine precisely $N(n_\cPqb^{\rm
  gen},n_{\rm q}^{\rm gen}, \scalht)$, $\epsilon$, and $m$ 
independently of $n_\cPqb$, which means that event yields for a
given b-quark jet multiplicity 
can be predicted with a higher statistical precision than obtained
directly from simulation. A precise determination of $m$ is particularly
important for events with $n_\cPqb \geq 3$, which occurs in the SM because of the
presence of mistagged jets in the event. In this case, the largest
background is \ttbar, with two correctly tagged b-quark jets and an
additional mistagged jet.

The magnitudes of the transfer factors are dependent on the control
sample and independent of the b-quark jet multiplicity, within
statistical uncertainties. For the \gj sample, the factors are also
independent of \scalht with values of approximately 0.4. For the \mj
and \mmj control samples, for which the \alphat requirement is dropped
from the selection criteria in the region $\scalht > 375\GeV$, the
factors decrease smoothly with increasing HT and are in the ranges 0.2
to 0.05 and 2 to 0.33, respectively. This variation arises from W +
jets and Z + jets events in the signal region, for which the
efficiency of the $\alphat > 0.55$ requirement is dependent on
\scalht.

A systematic uncertainty is assigned to each transfer factor to
account for theoretical uncertainties~\cite{Bern:2011pa} and also for
limitations in the simulation modelling of event
kinematics~\cite{RA1Paper2011}. The magnitudes of the uncertainties
are determined from a representative set of closure tests in data, in
which yields from one of the three independent control samples, along
with the corresponding transfer factors obtained from simulation, are
used to predict the yields in another control sample, following the
same prescription defined in Eq.~(\ref{equ:pred-method}). Hence, the
closure tests provide a consistency check between the predicted and
observed yields in the data control samples, from which the validity
of the method and the transfer factors can be established.

\begin{figure}[ht]
  \begin{center}
    \includegraphics[width=0.75\textwidth]{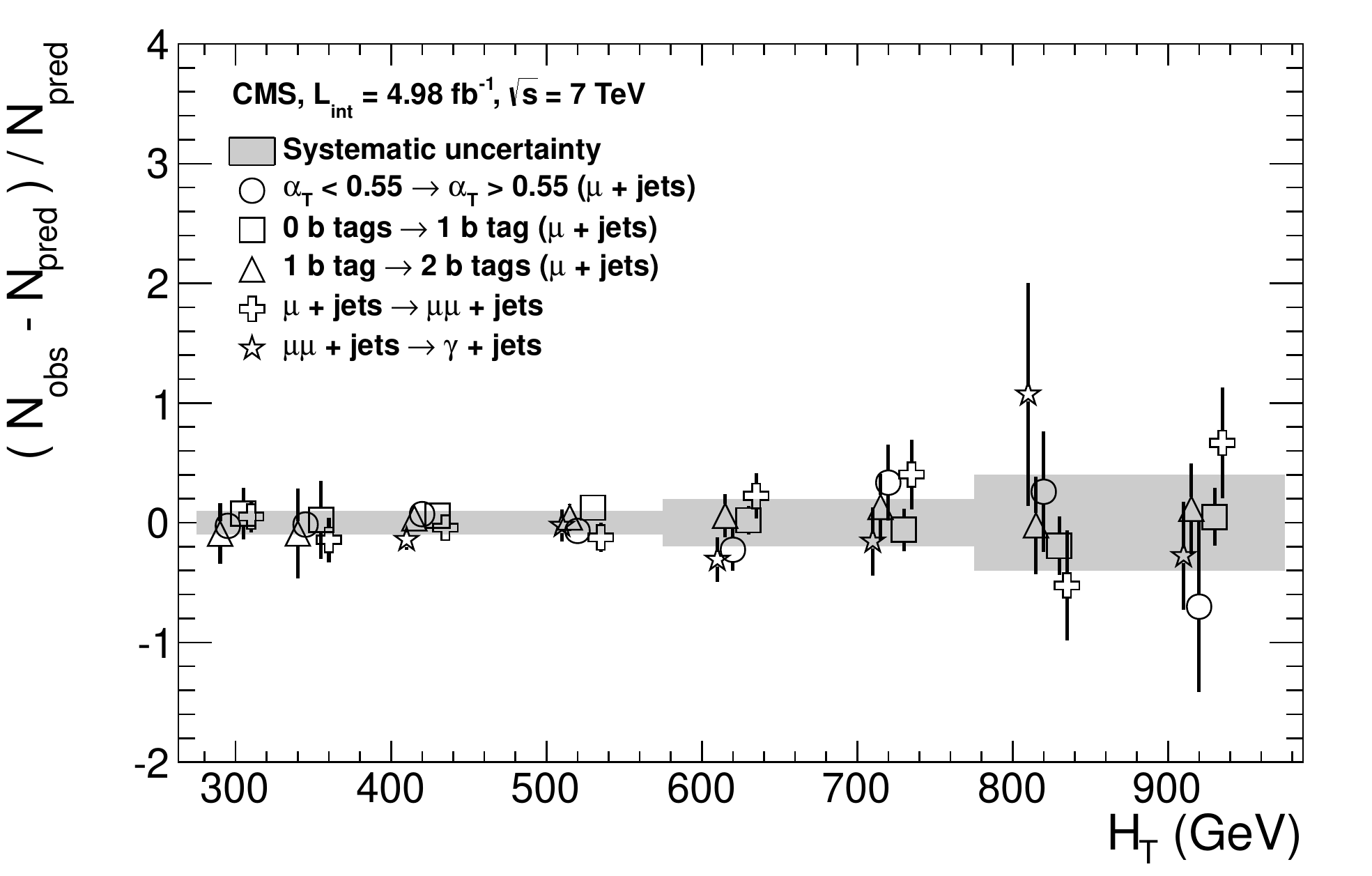}
    \caption{A set of five closure tests, described in the text, that
      use the three data control samples to probe key ingredients of
      the simulation modelling of the SM backgrounds, as a function of
      \scalht. Error bars represent statistical uncertainties
      only. The shaded bands represent the \scalht-dependent systematic
      uncertainties assigned to the transfer factors.  }
    \label{fig:closure-summary}
  \end{center}
\end{figure}

A set of five closure tests use the three data control samples to
probe key ingredients of the simulation modelling of the SM
backgrounds with genuine \met as a function of \scalht, as shown in
Fig.~\ref{fig:closure-summary}. The first three closure tests are
carried out within the $\mu$ + jets sample, and probe the modelling of
the \alphat distribution in genuine \met events (circles), the
relative contributions of $\PW$ + jets and \ttbar events (squares),
and the modelling of the reconstruction of b-quark jets (triangles),
respectively. The fourth test (crosses), connecting the $\mu$ + jets
and $\mu\mu$ + jets control samples, addresses the modelling of the
relative contributions of $\cPZ$ + jets to the sum of both $\PW$ + jets and \ttbar events,
while the fifth test (stars) deals with the consistency between the
Z$\rightarrow\mu\mu$ + jets and $\gamma$ + jets samples.  All
individual closure tests demonstrate, within the statistical precision
of each test, that there are no significant biases inherent in the
transfer factors obtained from simulation. The level of closure
achieved in these tests is used to estimate the systematic
uncertainties that are assigned to the transfer factors, which are
determined for three regions $275 < \scalht < 575\gev$, $575 < \scalht
< 775\gev$, and $\scalht > 775\gev$ to be 10\%, 20\%, and 40\%,
respectively.

A further dedicated study to account for potential systematic effects
arising from the modelling of the reconstruction of b-quark jets in the
simulation has been performed. After correcting the efficiency and
mistagging rates of b-quark jets in simulation for residual differences as
measured in data, the corresponding uncertainties on these corrections
are propagated to the transfer factors and found to be at the
sub-percent level. In addition, several robustness tests are
performed, including treating c-quark jets as b-quark jets in the yield estimates
throughout, as well as ignoring the contribution from hadronic
$\tau$-lepton decays. These tests also demonstrate sub-percent effects
on the transfer factors, highlighting the insensitivity to
potential mismodelling in simulation. Hence, the \scalht-dependent
systematic uncertainties of 10\%, 20\%, and 40\% are used for all b-quark
jet multiplicities.

\section{Results\label{sec:results}}

A likelihood model of the observations in all four data samples is
used to obtain a consistent prediction of the SM background, and to
test for the presence of a variety of signal models.  It is written as
\begin{align}
  \label{likelihood}
  L_{\text{total}} = &
  \prod\limits_{n_\cPqb=0}^{2}{\left(L_\text{hadronic}^{n_\cPqb} \times
    L_{\mu+\text{jets}}^{n_\cPqb} \times L_{\mu\mu+\text{jets}}^{n_\cPqb}
    \times L_{\gamma + \text{jets}}^{n_\cPqb}\right)} \nonumber \\
  & \times L_\text{hadronic}^{\geq 3} \times L_{\mu+\text{jets}}^{\geq
    3} \quad,
\end{align}
where $L_\text{hadronic}^{n_\cPqb}$ describes the yields in the
eight \scalht bins of the signal region when exactly $n_\cPqb$
reconstructed b-quark jets are required.  In each bin of \scalht, the
observation is modelled as Poisson-distributed about the sum of a SM
expectation and a potential signal contribution.  The components of
this SM expectation are related to the expected yields in the control
samples via transfer factors derived from simulation, as described
in Section~\ref{sec:methodology}. Signal contributions in each of the
four data samples are considered, though the only significant
contribution occurs in the signal region and not the control samples. The systematic
uncertainties associated with the transfer factors are accounted for with
nuisance parameters, the measurements of which are treated as
normally-distributed.  Since for $n_\cPqb\geq3$ the dominant SM
background arises from top events, only the $\mu$ + jets control
sample is used in the likelihood to determine the total contribution
from all (non-multijet) SM backgrounds in the signal region.

In addition, any potential contribution from multijet background in
the signal region is accounted for by using the ratio of events
which result in a value of \alphat above and below some threshold
value for a given \scalht bin. The dependence of this ratio, \RaT, on
\scalht is modelled as a falling exponential function: $A_{n_\cPqb}\Pe^{-k \, H_\mathrm{T}}$~\cite{RA1Paper}.  A common
parameter $k$ is used for all four categories of b-quark jet multiplicity,
and is constrained via measurements in a multijet-enriched data
side-band satisfying the criteria $\scalht < 575\gev$ and $0.52 <
\alphat < 0.55$.
A further side-band, defined by inverting the \mht/\met requirement of
Ref.~\cite{RA1Paper}, is used to confirm that this method provides an
unbiased estimate of $k$ and to determine a systematic uncertainty.

In order to test the compatibility of the observed yields with the
expectations from SM processes only, signal contributions are fixed to
zero and the likelihood function is maximised over all parameters.
The maximum likelihood values of the multijet normalisation
parameters $A_{n_\cPqb}$ are found to be compatible with zero,
within uncertainties, confirming the hypothesis that the multijet
background is negligible after the final selection.  Further, the SM
expected yields obtained from an alternate fit, in which these
normalisation parameters are fixed to zero, agree well with those
obtained from the nominal fit.

\begin{figure}[htbp]
  \begin{center}
    \subfigure[Signal region]{\includegraphics[width=0.45\textwidth]{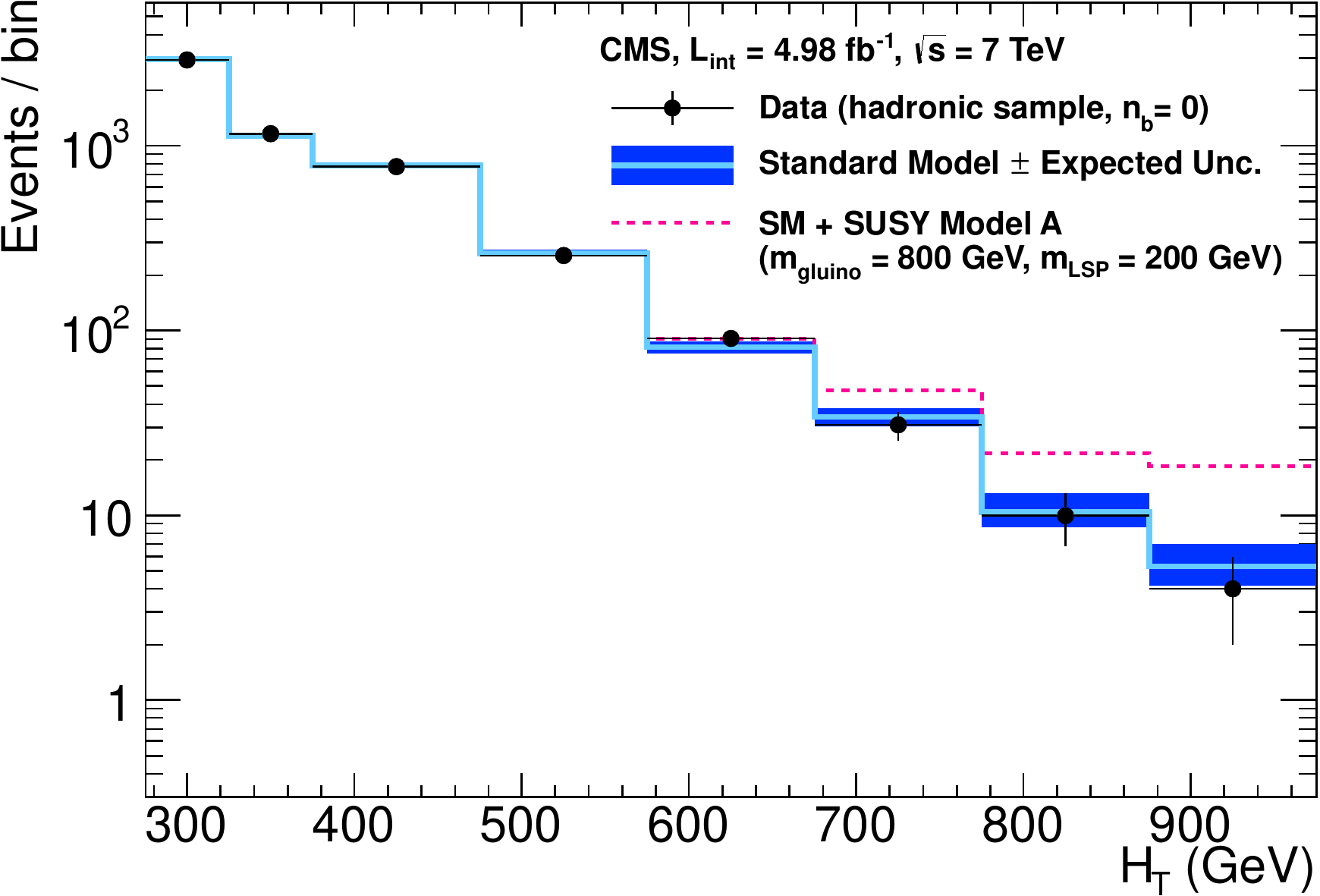}} \qquad
    \subfigure[$\mu$ + jets sample]{\includegraphics[width=0.45\textwidth]{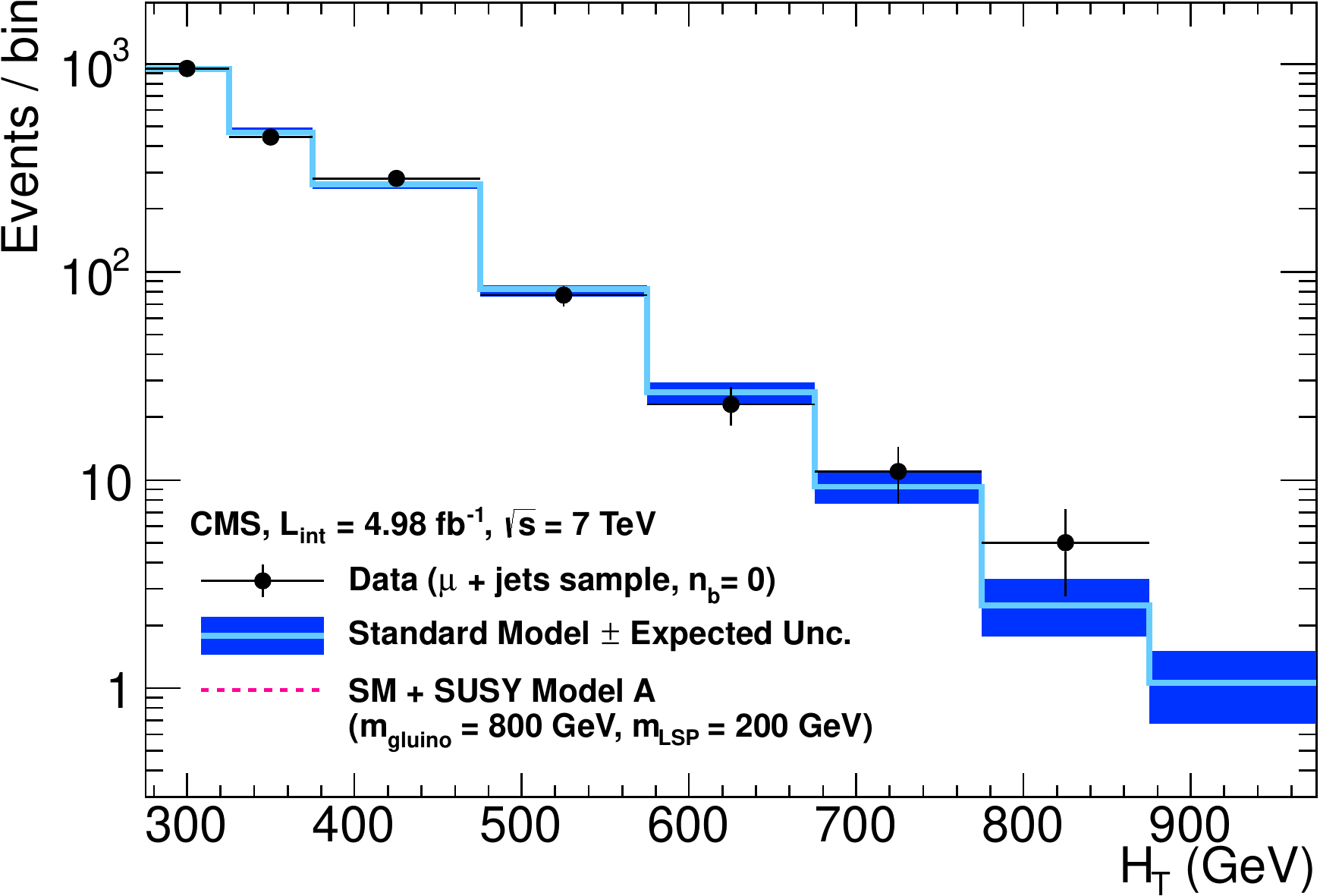}} \\
    \subfigure[$\mu\mu$ + jets sample]{\includegraphics[width=0.45\textwidth]{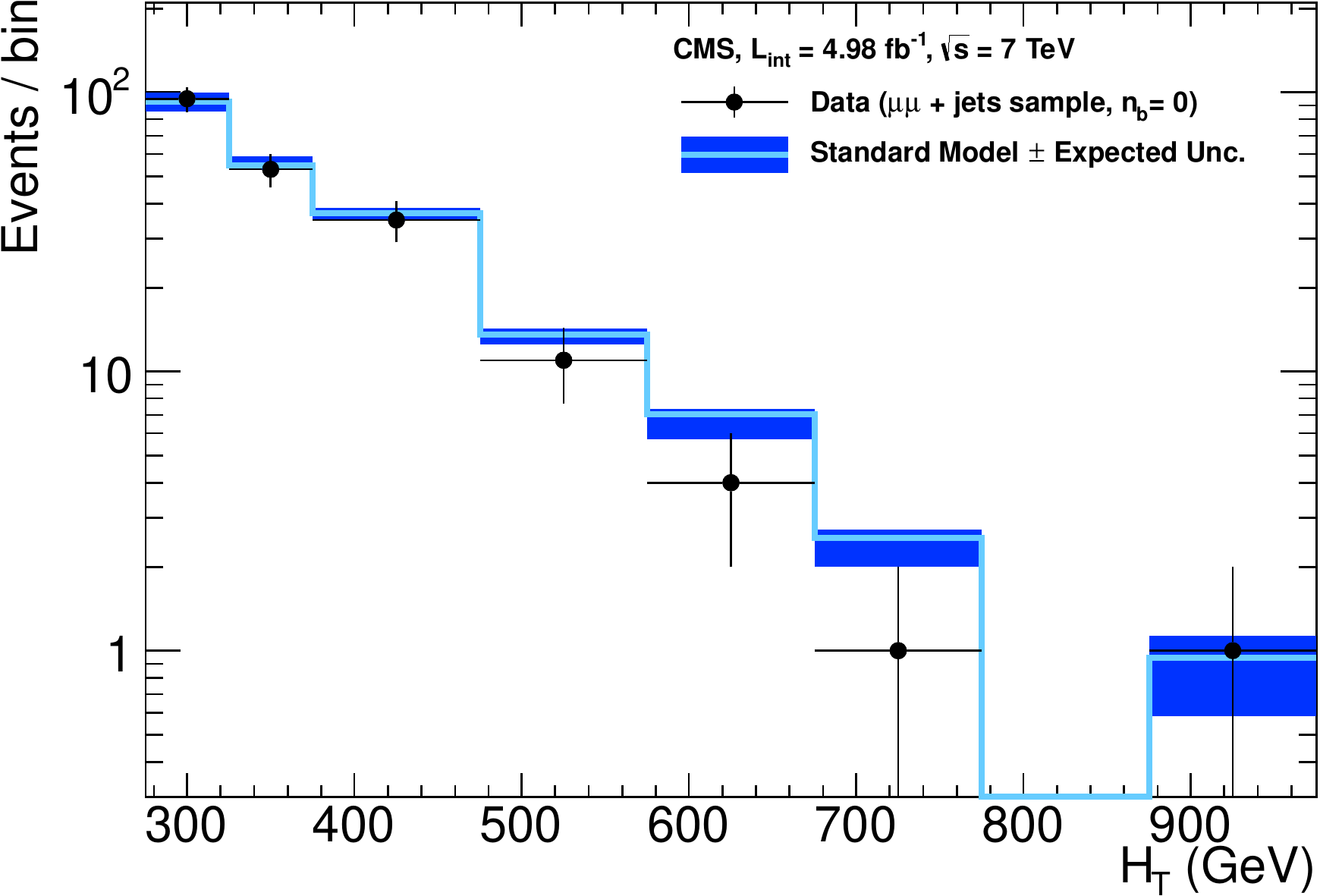}} \qquad
    \subfigure[$\gamma$ + jets sample]{\includegraphics[width=0.45\textwidth]{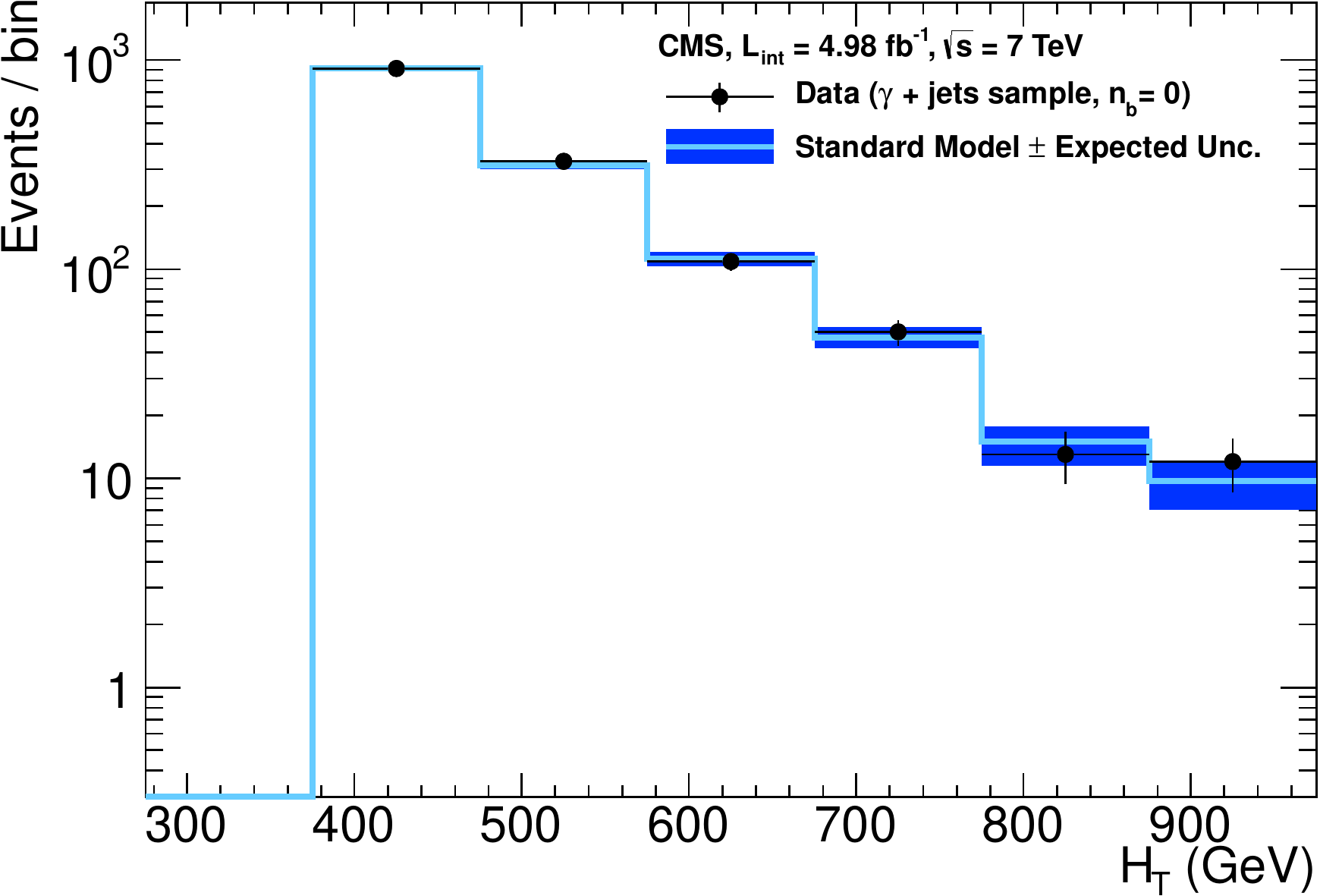}} \\
    \caption{\label{fig:best-fit-0-btag} Comparison of the observed
      yields and SM expectations given by the simultaneous fit in bins
      of \scalht for the (a) signal region, (b) \mj, (c) \mmj, and (d) \gj
      samples when requiring exactly zero reconstructed b-quark jets. The
      observed event yields in data (black dots) and the expectations
      and their uncertainties, as determined by the simultaneous fit,
      for all SM processes (light blue solid line with dark blue
      bands) are shown.
      For illustrative purposes only, the signal expectation (magenta
      dashed line) in the signal region for the simplified model $A$ (defined in Section~\ref{sec:sms}) with $m_{\gl} =
      800\gev$ and $m_\mathrm{LSP} = 200\gev$ is superimposed on the SM
      background expectation.
    }
  \end{center}
\end{figure}

\begin{figure}[htbp]
  \begin{center}
    \subfigure[Signal region]{\includegraphics[width=0.45\textwidth]{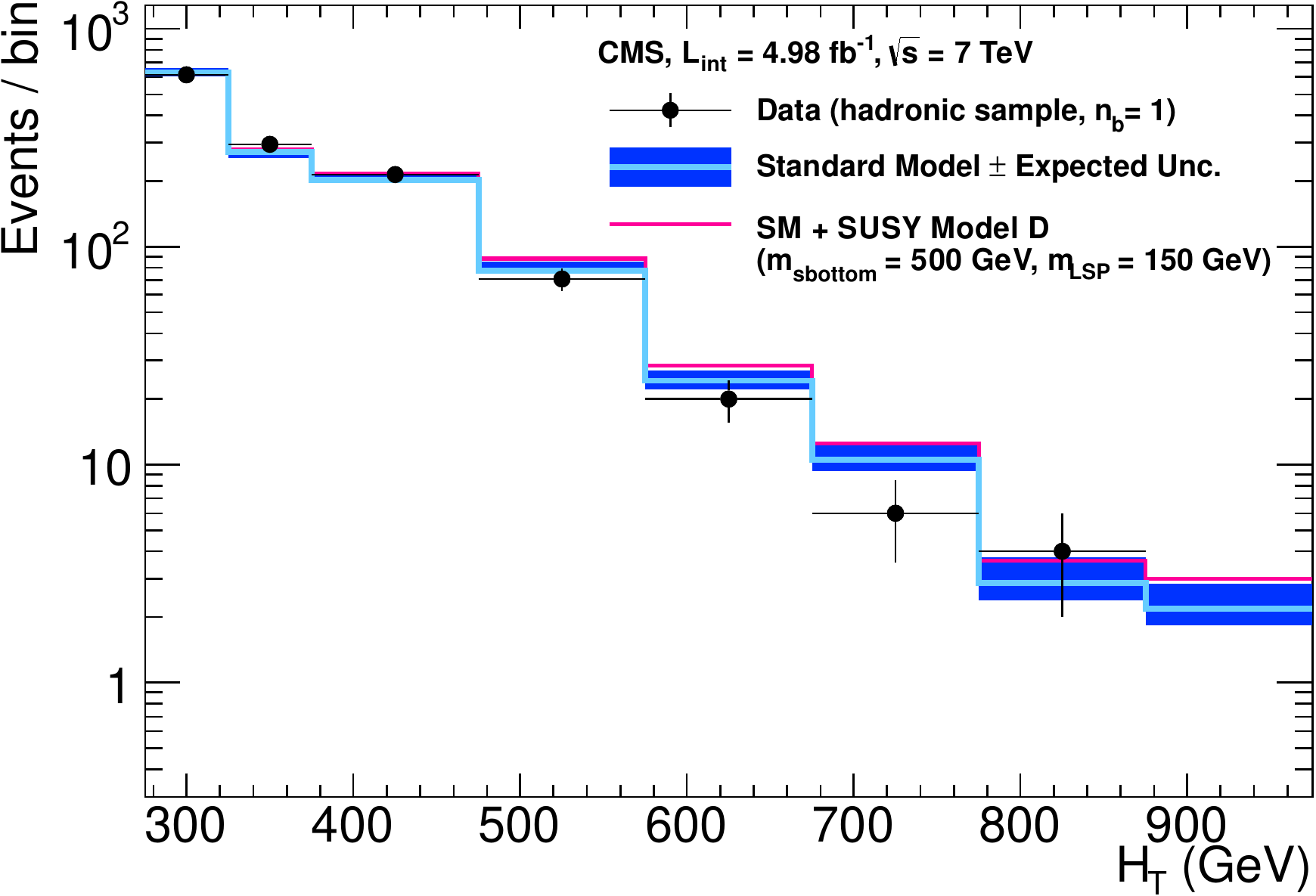}} \qquad
    \subfigure[$\mu$ + jets sample]{\includegraphics[width=0.45\textwidth]{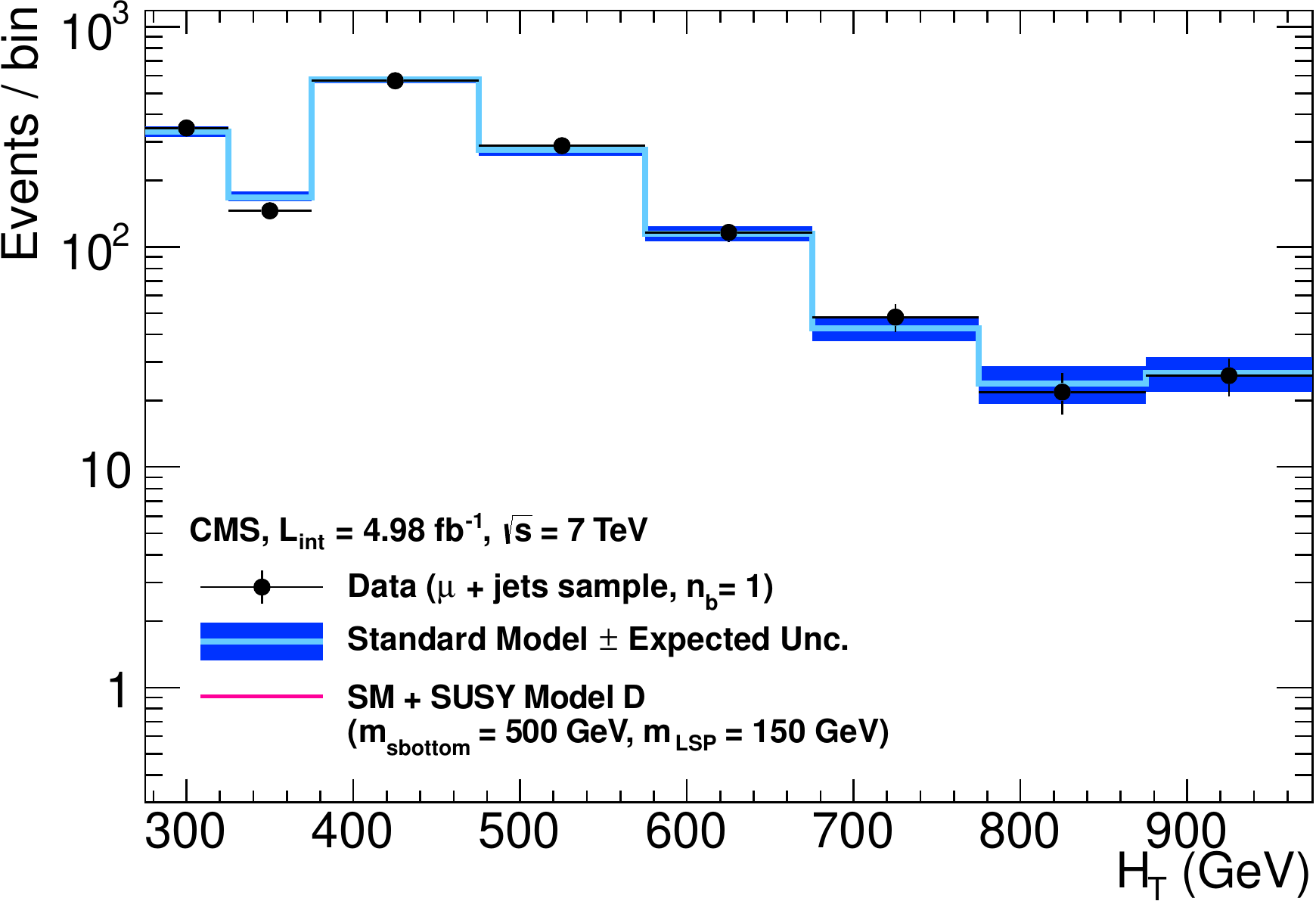}} \\
    \subfigure[$\mu\mu$ + jets sample]{\includegraphics[width=0.45\textwidth]{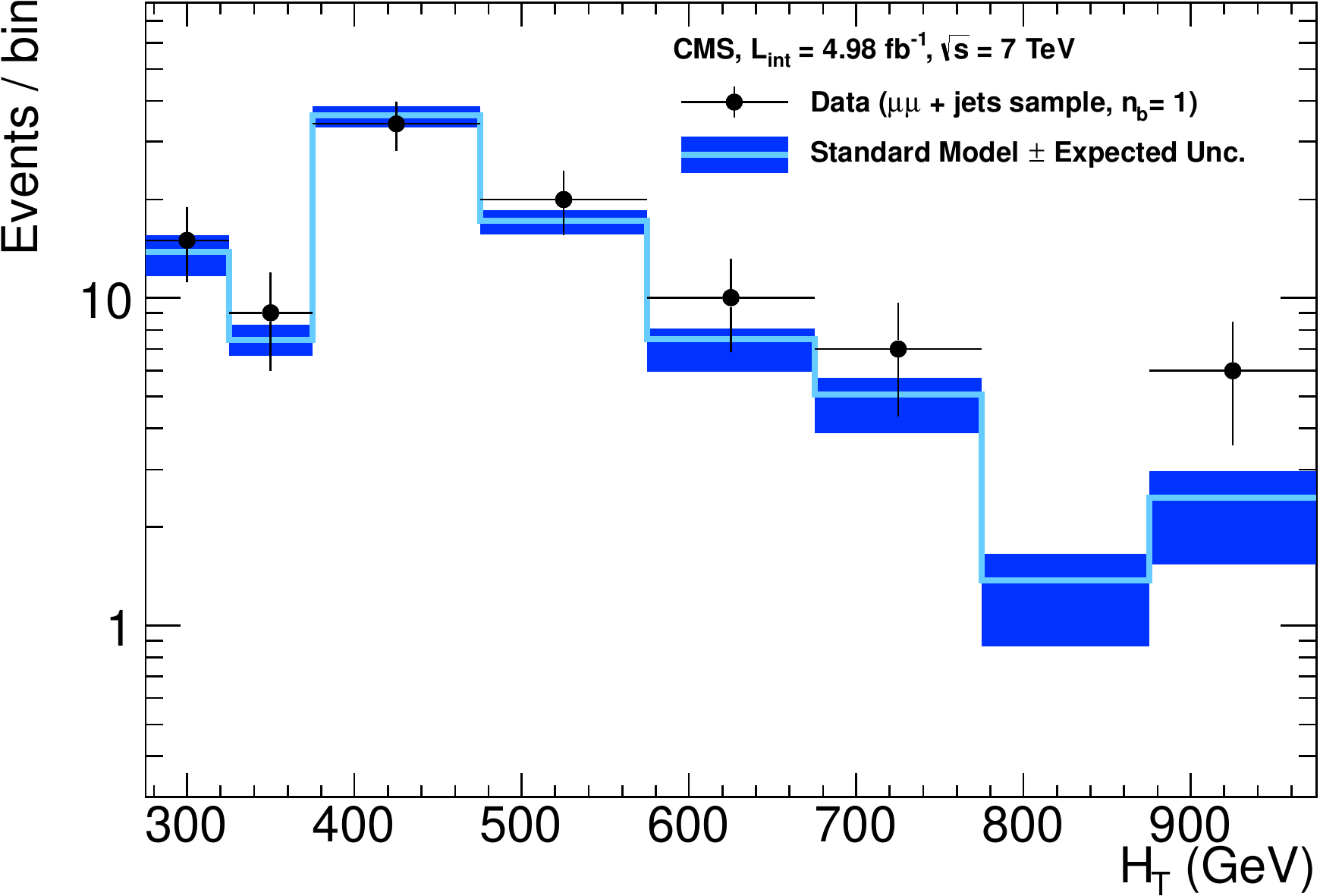}} \qquad
    \subfigure[$\gamma$ + jets sample]{\includegraphics[width=0.45\textwidth]{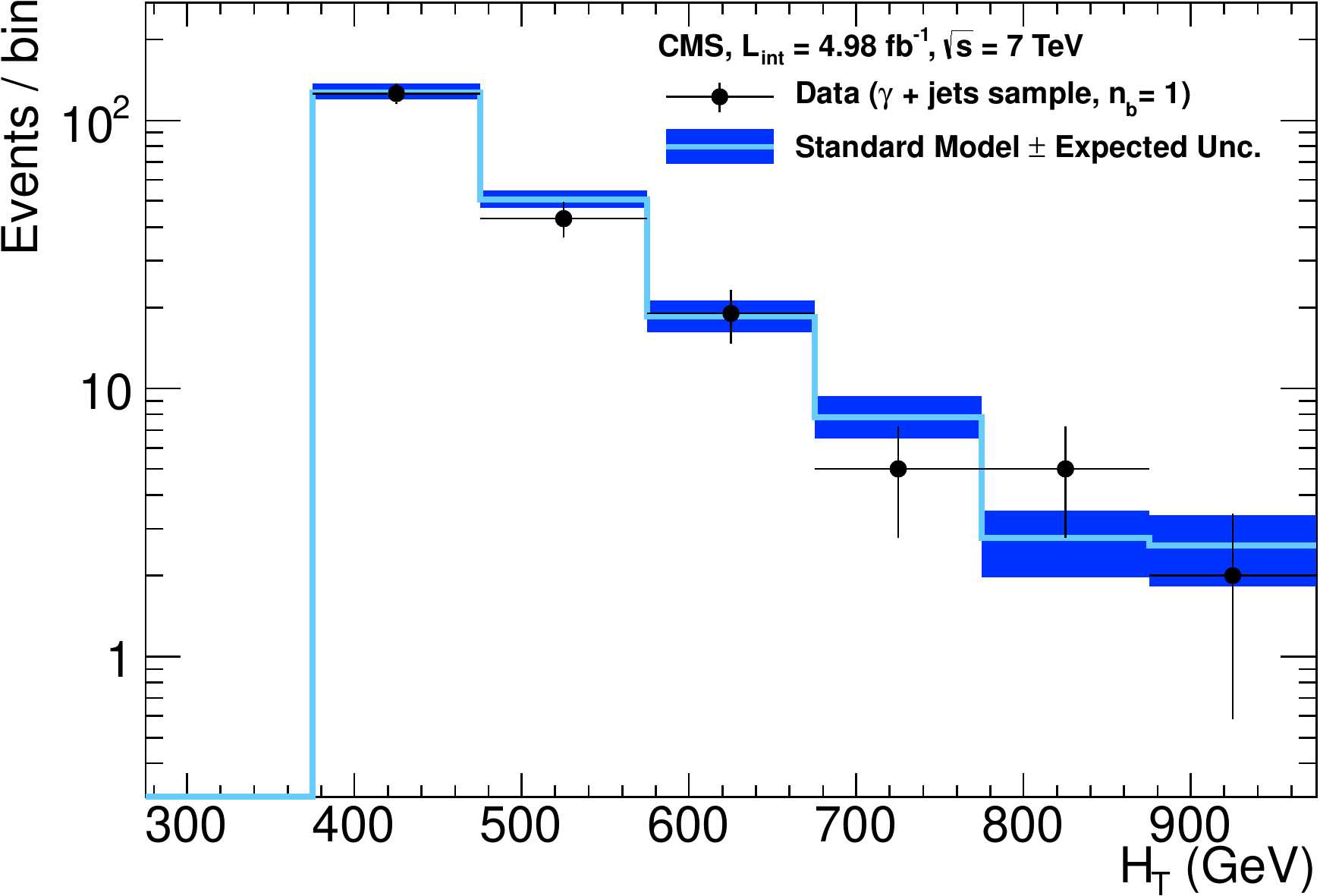}} \\
    \caption{\label{fig:best-fit-1-btag} Comparison of the observed
      yields and SM expectations given by the simultaneous fit in bins
      of \scalht for the (a) signal region, (b) \mj, (c) \mmj, and (d) \gj
      samples. Same as Fig.~\ref{fig:best-fit-0-btag}, except
      requiring exactly one reconstructed b-quark jet. The observed event
      yields in data (black dots) and the expectations and their
      uncertainties, as determined by the simultaneous fit, for all SM
      processes (light blue solid line with dark blue bands) are
      shown.
      For illustrative purposes only, the signal expectation (magenta
      solid line) in the signal region for the simplified model $D$ (defined in Section~\ref{sec:sms}) with $m_{\gl} =
      500\gev$ and $m_\mathrm{LSP} = 150\gev$ is superimposed on the SM
      background expectation.
    }
  \end{center}
\end{figure}

\begin{figure}[htbp]
  \begin{center}
    \subfigure[Signal region]{\includegraphics[width=0.45\textwidth]{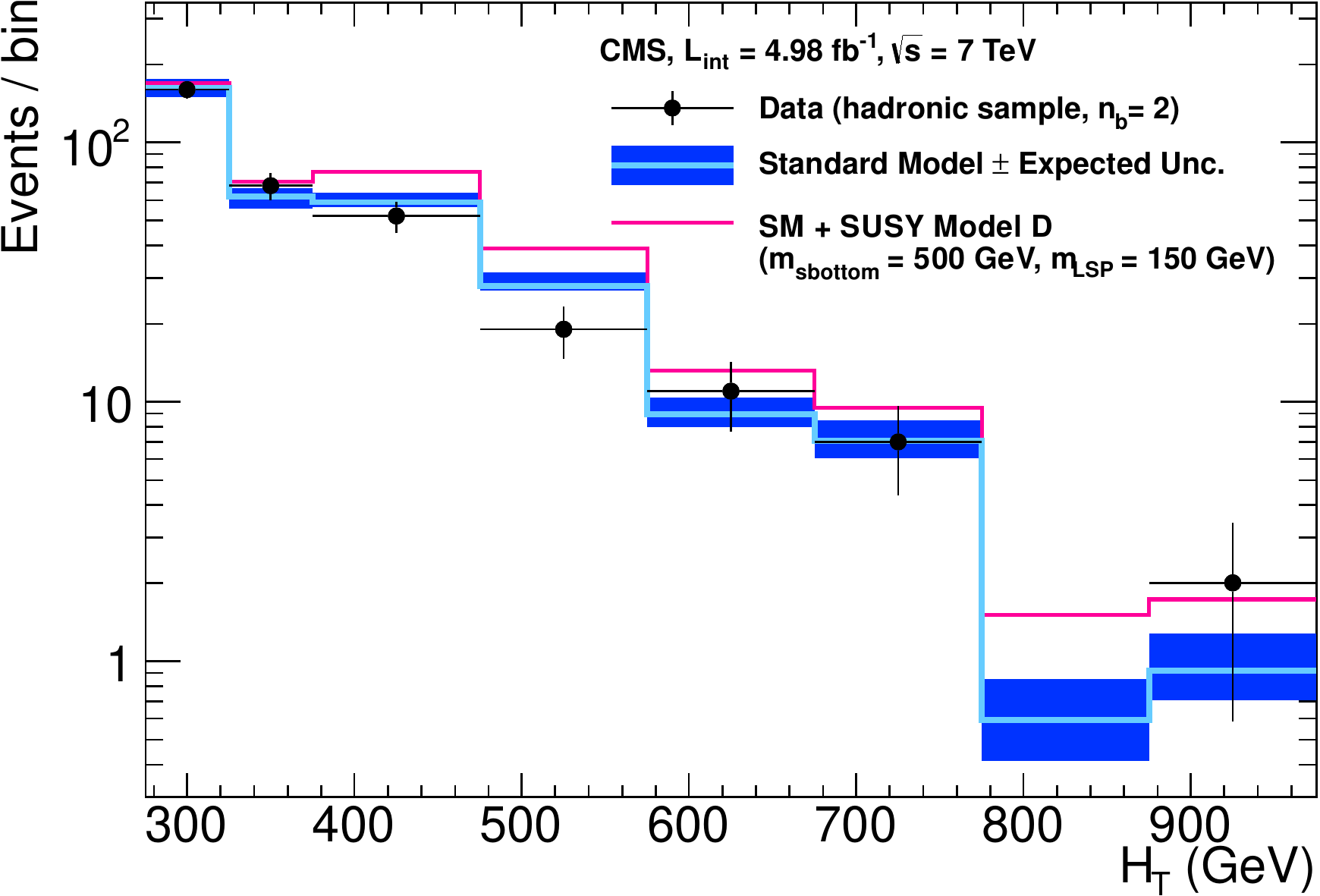}} \qquad
    \subfigure[$\mu$ + jets sample]{\includegraphics[width=0.45\textwidth]{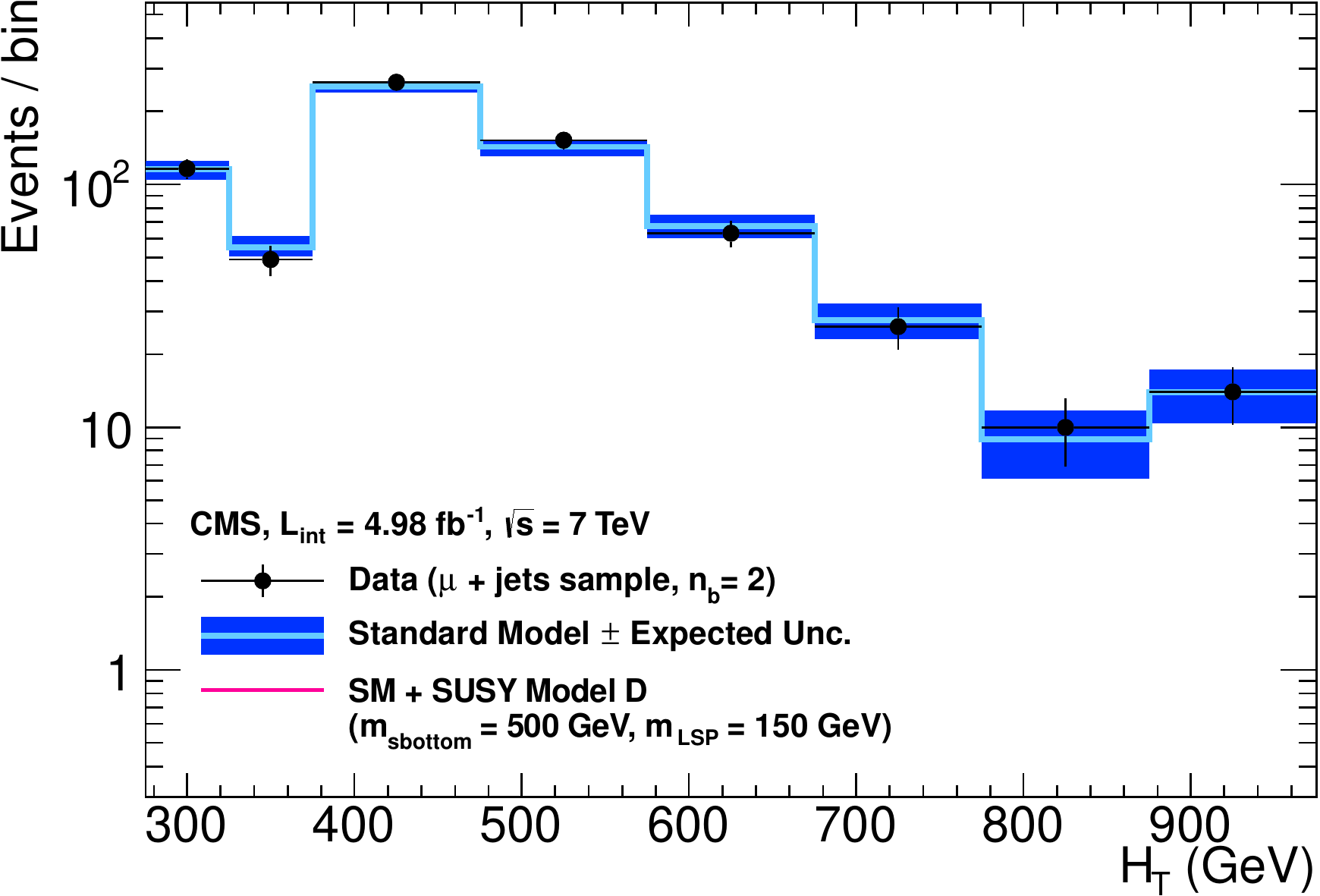}} \\
    \subfigure[$\mu\mu$ + jets sample]{\includegraphics[width=0.45\textwidth]{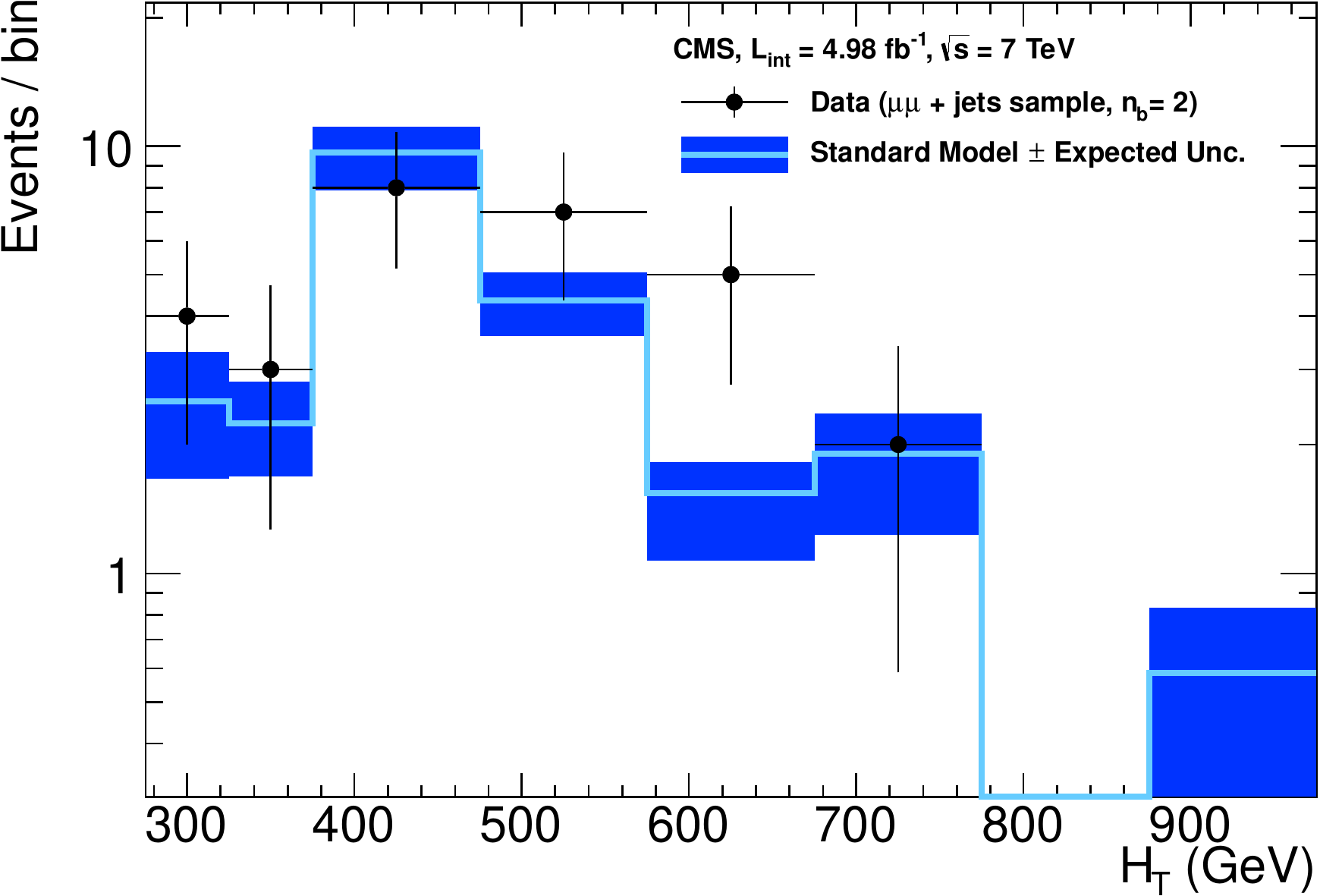}} \qquad
    \subfigure[$\gamma$ + jets sample]{\includegraphics[width=0.45\textwidth]{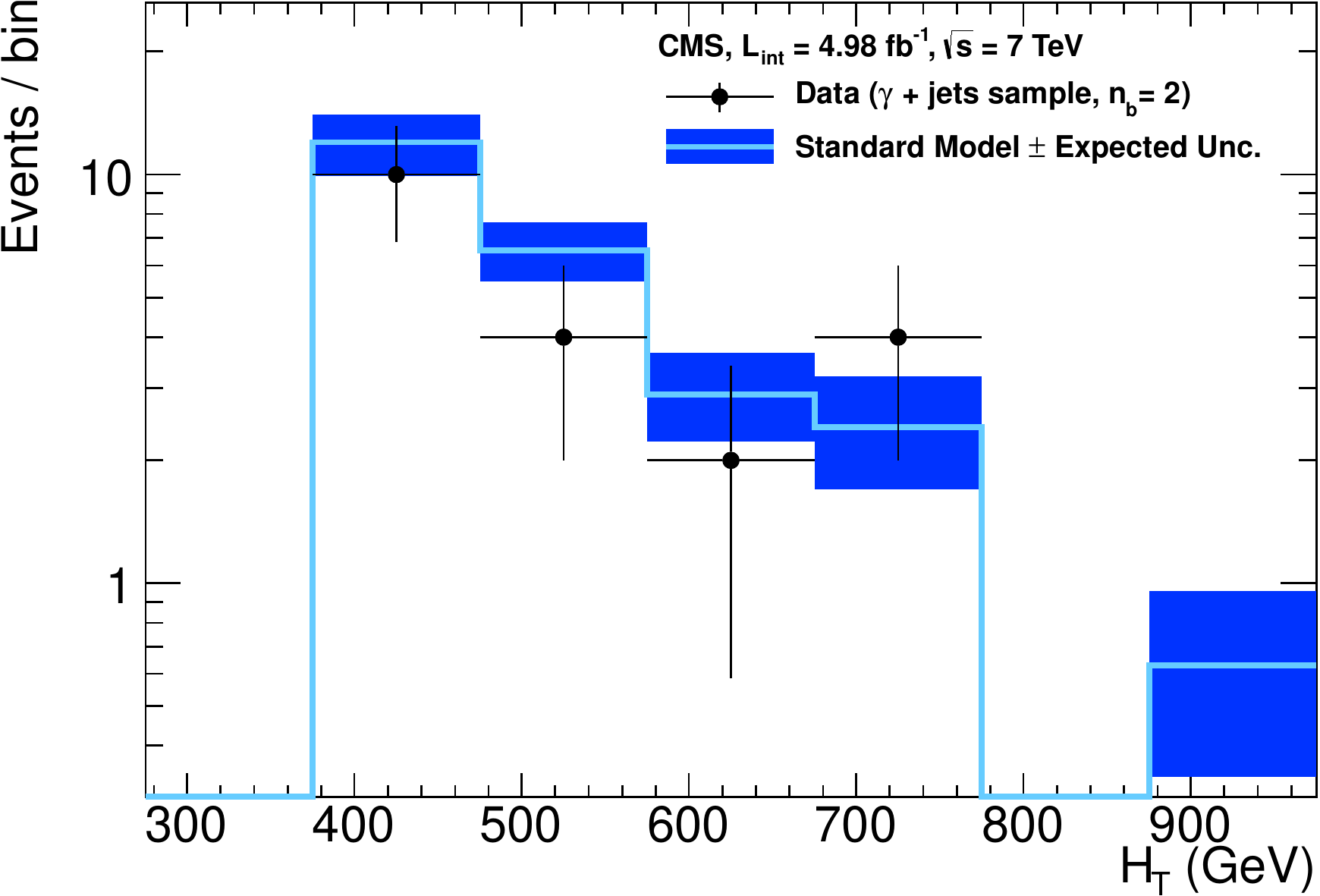}} \\
    \caption{\label{fig:best-fit-2-btag} Comparison of the observed
      yields and SM expectations given by the simultaneous fit in bins
      of \scalht for the (a) signal region, (b) \mj, (c) \mmj, and (d) \gj
      samples. Same as Fig.~\ref{fig:best-fit-0-btag}, except
      requiring exactly two reconstructed b-quark jets. The observed event
      yields in data (black dots) and the expectations and their
      uncertainties, as determined by the simultaneous fit, for all SM
      processes (light blue solid line with dark blue bands) are
      shown.
      For illustrative purposes only, the signal expectation (magenta
      solid line) in the signal region for the simplified model $D$ (defined in Section~\ref{sec:sms}) with $m_{\gl} =
      500\gev$ and $m_\mathrm{LSP} = 150\gev$ is superimposed on the SM
      background expectation.
    }
  \end{center}
\end{figure}

\begin{figure}[htbp]
  \begin{center}
    \subfigure[Signal region]{\includegraphics[width=0.45\textwidth]{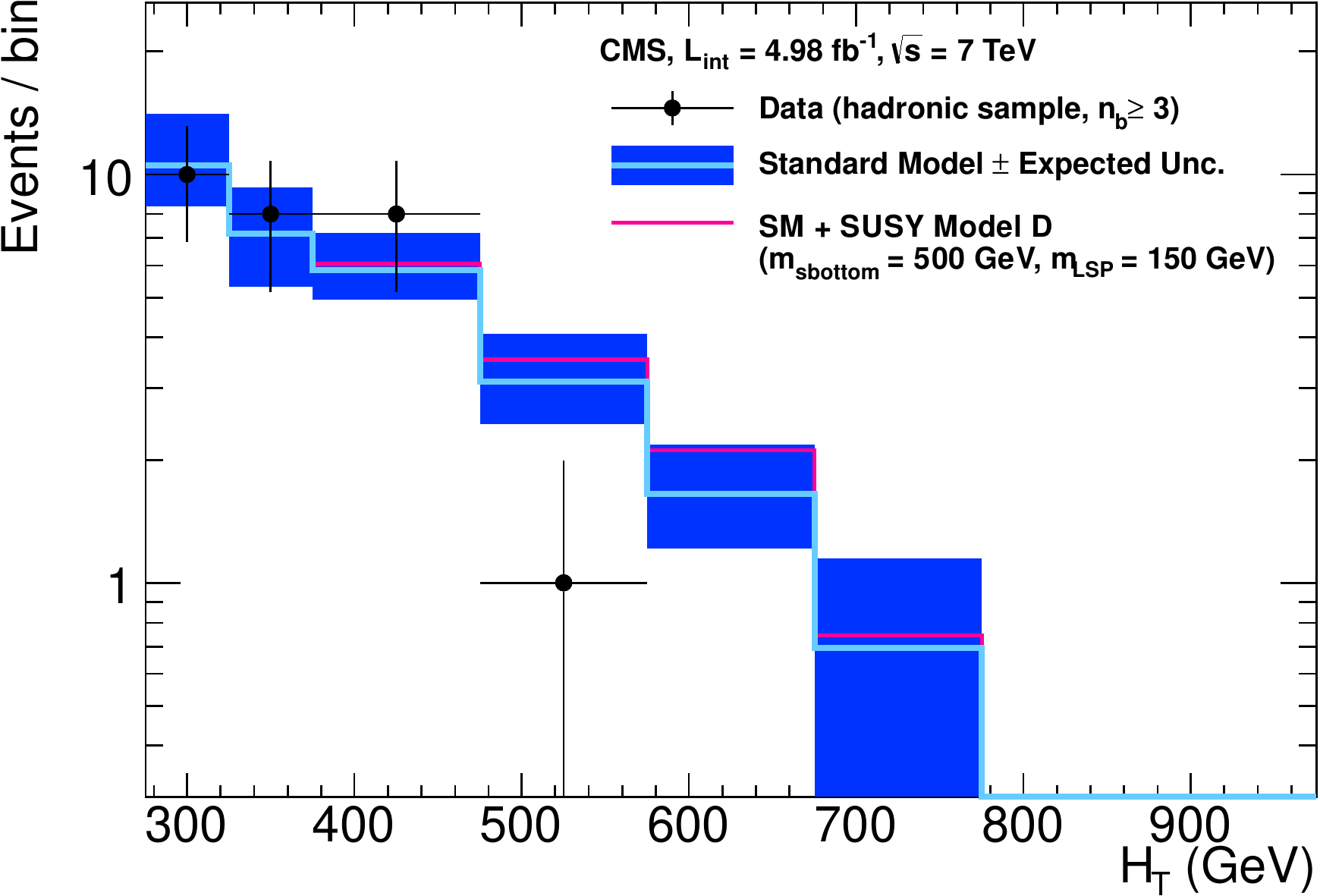}} \qquad
    \subfigure[$\mu$ + jets sample]{\includegraphics[width=0.45\textwidth]{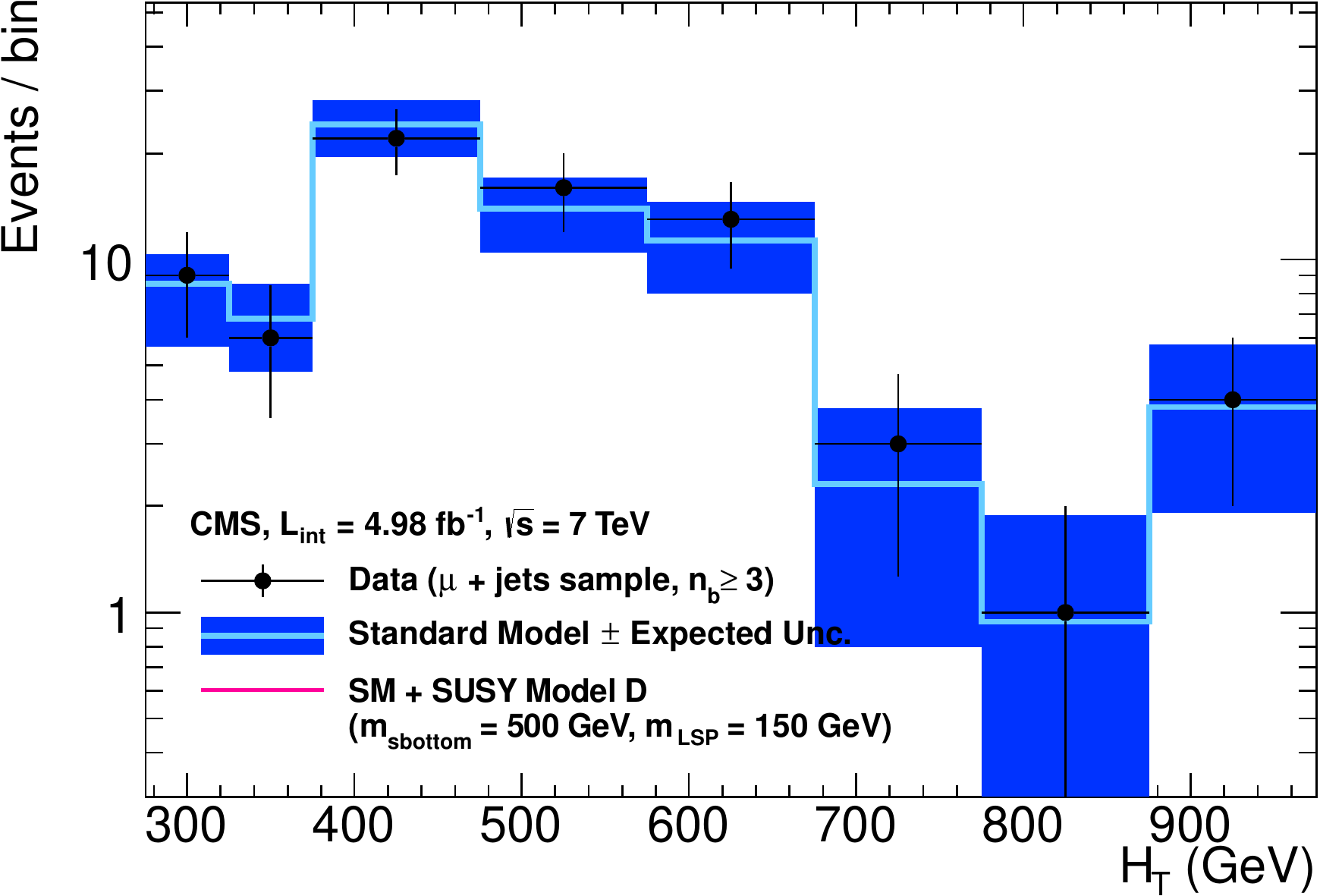}} \\
    \caption{\label{fig:best-fit-3-btag} Comparison of the observed
      yields and SM expectations given by the simultaneous fit in bins
      of \scalht for the (a) signal region and (b) \mj samples. Same as
      Fig.~\ref{fig:best-fit-0-btag}, except requiring at least three
      reconstructed b-quark jets. The observed event yields in data (black
      dots) and the expectations and their uncertainties, as
      determined by the simultaneous fit, for all SM processes (light
      blue solid line with dark blue bands) are shown.
      For illustrative purposes only, the signal expectation (magenta
      solid line) in the signal region for the simplified model $D$ (defined in Section~\ref{sec:sms}) with $m_{\gl} =
      500\gev$ and $m_\mathrm{LSP} = 150\gev$ is superimposed on the SM
      background expectation.
    }
  \end{center}
\end{figure}

The signal region data yields, as well as the SM expectations
obtained from the simultaneous fit across all samples, are shown in
Table~\ref{fit-summary}.  A comparison of the observed yields and the
SM expectations in bins of \scalht for events with exactly zero, one,
two, and at least three reconstructed b-quark jets are shown in
Figures~\ref{fig:best-fit-0-btag}, \ref{fig:best-fit-1-btag},
\ref{fig:best-fit-2-btag}, and \ref{fig:best-fit-3-btag},
respectively, for the signal region and the three control
samples.  In all four categories of b-quark jet multiplicity, the
samples are well described by the SM hypothesis.  In particular, no
significant excess above the SM expectation is observed in the
signal region.

\begin{table}[tbh]
\topcaption{Comparison of the observed yields in the different \scalht
  and b-quark jet multiplicity bins for the signal region with the SM
  expectations and combined statistical and systematic uncertainties
  given by the simultaneous fit.}
\label{fit-summary}
\footnotesize
\setlength{\extrarowheight}{2.5pt}
\centering
\begin{tabular}{ lllllllll }
\hline
\multicolumn{1}{@{}r}{\textbackslash\scalht\,[\GeVns{}]}        & 275--325             & 325--375             & 375--475             & 475--575             & 575--675             & 675--775             & 775--875             & $>$875       \\
 \# b-quark jets\textbackslash &&&&&&&&\\
\hline
0 (SM)          & $2933^{+56}_{-52}$   & $1139^{+17}_{-40}$   & $783^{+17}_{-27}$    & $261^{+14}_{-8}$     & $81.5^{+6.5}_{-6.5}$ & $34.2^{+4.0}_{-3.8}$ & $10.4^{+2.8}_{-1.8}$ & $5.3^{+1.7}_{-1.1}$ \\
0 (Data)        & $2919$               & $1166$               & $769$                & $255$                & $91$                 & $31$                 & $10$                 & $4$                 \\
\hline
1 (SM)          & $630^{+26}_{-25}$    & $271^{+10}_{-16}$    & $202^{+10}_{-6}$     & $78.0^{+6.9}_{-1.9}$ & $24.2^{+2.9}_{-2.0}$ & $10.6^{+1.7}_{-1.3}$ & $2.9^{+0.9}_{-0.5}$  & $2.2^{+0.7}_{-0.4}$ \\
1 (Data)        & $614$                & $294$                & $214$                & $71$                 & $20$                 & $6$                  & $4$                  & $0$                 \\
\hline
2 (SM)          & $162^{+13}_{-12}$    & $61.8^{+4.8}_{-6.3}$ & $58.8^{+4.8}_{-2.6}$ & $28.0^{+3.5}_{-1.1}$ & $9.0^{+1.4}_{-1.0}$  & $7.1^{+1.4}_{-1.0}$  & $0.6^{+0.3}_{-0.2}$  & $0.9^{+0.4}_{-0.2}$ \\
2 (Data)        & $160$                & $68$                 & $52$                 & $19$                 & $11$                 & $7$                  & $0$                  & $2$                 \\
\hline
$\geq$3  (SM)   & $10.5^{+3.5}_{-2.2}$ & $7.1^{+2.2}_{-1.8}$  & $5.8^{+1.4}_{-0.9}$  & $3.1^{+1.0}_{-0.7}$  & $1.7^{+0.5}_{-0.4}$  & $0.7^{+0.5}_{-0.4}$  & $0.1^{+0.1}_{-0.1}$  & $0.2^{+0.1}_{-0.1}$ \\
$\geq$3 (Data) & $10$                 & $8$                  & $8$                  & $1$                  & $0$                  & $0$                  & $0$                  & $0$                 \\
\hline
\end{tabular}
\end{table}

\section{Interpretation of the results\label{sec:interpretation}}

Limits are set in the parameter space of the CMSSM and in a set of
simplified models that characterise both third-generation squark
production and compressed SUSY spectra scenarios.  The \cls
method~\cite{read,junk} is used to compute the limits, with the 
one-sided profile likelihood ratio as the
test statistic~\cite{Cowan:2010js}.  The sampling distributions for
the test statistic are built by generating pseudo-data from the
likelihood function, using the respective maximum-likelihood values of
the nuisance parameters under the background-only and
signal-plus-background hypotheses.

Events samples for the CMSSM and simplified models are generated at
leading order with \PYTHIA6.4~\cite{pythia}. Inclusive,
process-dependent, next-to-leading order calculations with
next-to-leading logarithmic corrections~\cite{ref:susy-nlo-nll-1, ref:susy-nlo-nll-2, ref:susy-nlo-nll-3, ref:susy-nlo-nll-4, ref:susy-nlo-nll-5} (NLO+NLL)
of SUSY production cross sections are obtained with the program
\PROSPINO~\cite{Beenakker:1996ch} and CTEQ6M~\cite{Pumplin:2002vw}
parton distribution functions. The simulated signal events include
multiple interactions per LHC bunch crossing (pileup) with the
distribution of reconstructed vertices that match the one observed in
data.

\subsection{Interpretation in the CMSSM}

The CMSSM is described by the following five parameters:
the universal scalar and gaugino mass parameters, $m_0$ and $m_{1/2}$;
the universal trilinear soft SUSY-breaking parameter, $A_0$; the ratio
of the vacuum expectation values of the two Higgs doublets,
$\tan\beta$; and the sign of the Higgs mixing parameter, $\mu$.
At each point in the parameter space of the CMSSM, the SUSY particle
spectrum is calculated with \textsc{SoftSUSY} \cite{Allanach:2001kg}.
Experimental uncertainties on the SM background prediction
(10--40\%), the luminosity measurement (2.2\%)~\cite{ref:lumi}, and
the total selection efficiency times acceptance for the considered
signal model (16\%) are included in the calculation of the limit. The
dominant sources of uncertainty on the signal efficiency times
acceptance are derived from systematic variations of parton
distribution functions, and corrections applied to jet energies and
b-tagging efficiency and mistag rates.

\begin{figure}[htbp]
  \begin{center}
    \includegraphics[width=0.7\textwidth]{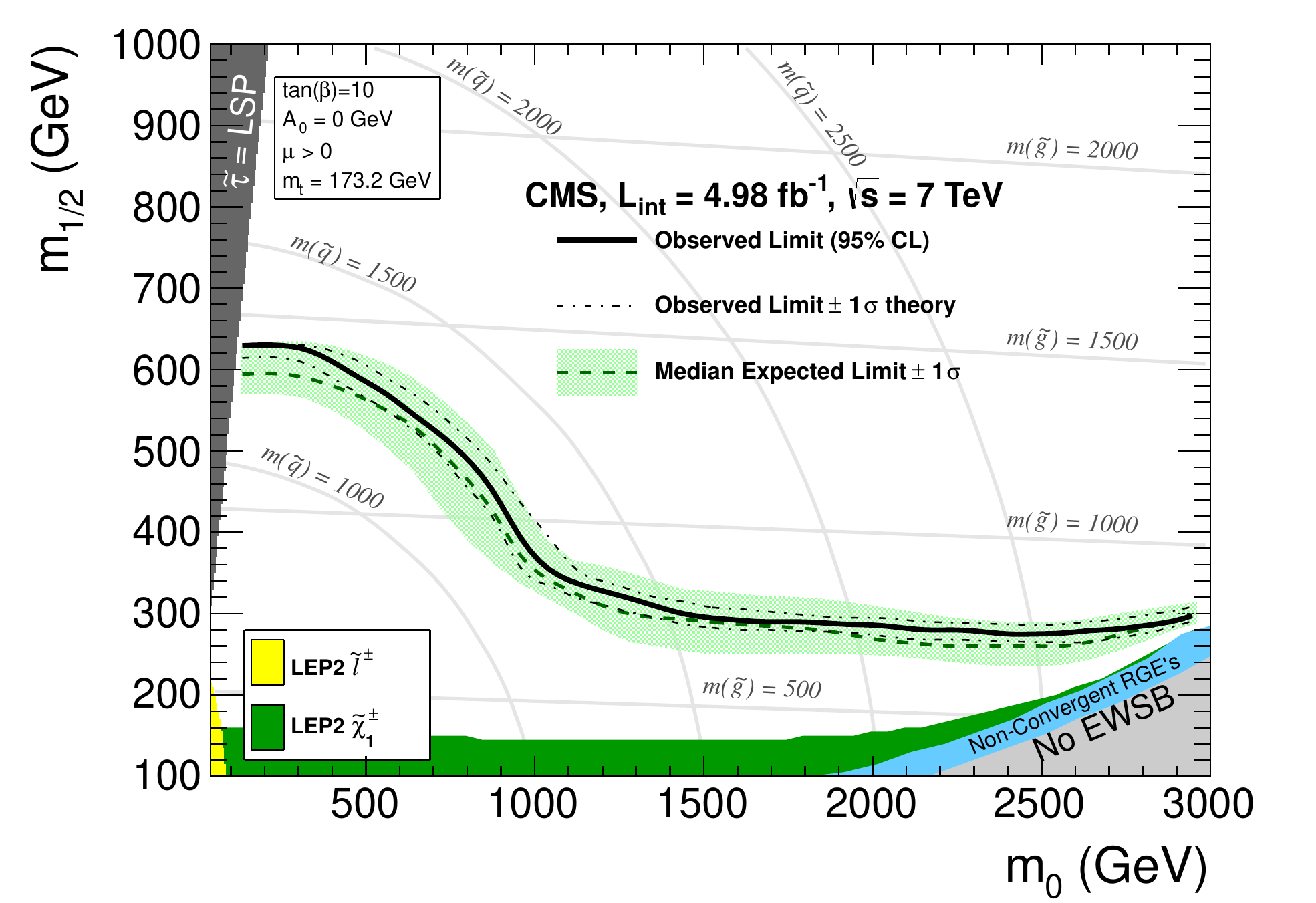}
    \caption{\label{fig:cmssm-limit} Exclusion contours at 95\% CL in
      the CMSSM ($m_0, m_{1/2}$) plane ($\tan \beta = 10, A_0 = 0, \mu
      > 0$) calculated with NLO+NLL SUSY production cross sections and
      the \cls method. The solid black line indicates the observed
      exclusion region. The dotted-dashed black lines represent the
      observed excluded region when varying the production cross
      section by its theoretical uncertainty. The expected median
      exclusion region (green dashed line) $\pm 1 \sigma$ (green band)
      are also shown. The CMSSM template is taken from
      Ref.~\cite{cmssm-template}. }
  \end{center}
\end{figure}

Figure~\ref{fig:cmssm-limit} shows the observed and expected exclusion
limits at 95\% confidence level (CL) in the $(m_0 , m_{1/2})$ plane
for $\tan \beta = 10$ and $A_0 = 0\GeV$, calculated with the NLO+NLL
SUSY production cross section. For this choice of parameter values,
squark masses below $1250\GeV$ are excluded at 95\% CL, as are gluino
masses below the same value for the region $m_0 < 600\GeV$.  In the
region $600 < m_0 < 3000\GeV$, gluino masses below $700\GeV$ are
excluded, while the squark mass in the excluded models varies in the
range 1250--2500\GeV, depending on the value of $m_0$. The mass
limits are determined conservatively from the observed exclusion based
on the theoretical production cross section minus $1\sigma$
uncertainty~\cite{ref:susy-theory}.

\subsection{Interpretation with simplified models\label{sec:sms}}

The data observations are also interpreted using simplified models
that characterise third-generation squark production and compressed
spectra scenarios, where the mass difference between the primary
produced sparticle (e.g. a squark or a gluino) and the LSP is rather
small. The production and decay modes of the models under
consideration are summarised in Table~\ref{tab:sms}. The simplified
models $A$ and $B$ are used to characterise the pair
production of gluinos and first- or second-generation squarks,
respectively, depending on their mass as well as on the LSP
mass. Simplified models $C$ to $F$ describe various
production and decay mechanisms in the context of third-generation
squarks.

Experimental uncertainties on the SM background predictions
(10--40)\%, the luminosity measurement (2.2\%), and the total
acceptance times efficiency of the selection for the considered signal
model (12\%$-$18\%) are included in the calculation of the
limit. Signal efficiency in the kinematic region defined
by $0 < m_{\sGlu(\sQua)} - m_{\textrm{LSP}} < 175\gev$ or
$m_{\sGlu(\sQua)} < 300\gev$ is due in part to the presence of
initial-state radiation. Given the large associated uncertainties, no
interpretation is provided for this kinematic region. In the case of
model $E$, for which pair-produced gluinos decay to \ttbar
pairs and the LSP, the region $0 < m_{\sGlu} - m_{\textrm{LSP}} <
400\gev$ is not considered.

\begin{table}[tbh]
  \topcaption{The first three columns define the production and decay
    modes for various simplified models. The last two columns
    indicate the search sensitivity for these models, where
    $m_{\sq(\sGlu)}^{\textrm{best}}$ and
    $m_{\textrm{LSP}}^{\textrm{best}}$ represent the largest mass
    beyond which no limit can be set for squarks/gluinos and the LSP,
    respectively. The exclusion range for $m_{\sq(\sGlu)}$ is bounded
    from below by the kinematic region considered for each simplified
    model, as defined in the text. The quoted estimates are determined
    conservatively from the observed exclusion based on the theoretical production cross
    section minus $1\sigma$ uncertainty. For model $C$, the search
    is at the threshold of sensitivity for the considered
    ($m_{\sQua},m_\mathrm{LSP}$) parameter space, as discussed in the text.
  }
  \label{tab:sms}
  \centering
  \begin{tabular}{ llccc }
    \hline
    Model & Production and decay modes & Figure & $m_{\sq(\sGlu)}^{\textrm{best}}$~(\GeVns{}) & $m_{\textrm{LSP}}^{\textrm{best}}$~(\GeVns{}) \\ [0.5ex]
    \hline
    $A$ &
    $\textrm{pp}\,\rightarrow\,\sGlu\sGlu\,\rightarrow\,\textrm{q}\bar{\textrm{q}}\chiz\textrm{q}\bar{\textrm{q}}\chiz$
    & \ref{fig:t1}  & $\approx$950 & $\approx$400 \\
    $B$ &
    $\textrm{pp}\,\rightarrow\,\sQua\sQua\,\rightarrow\,\textrm{q}\chiz\bar{\textrm{q}}\chiz$
    & \ref{fig:t2}  & $\approx$750 & $\approx$275 \\
    $C$ &
    $\textrm{pp}\,\rightarrow\,\sTop\sTop\,\rightarrow\,\textrm{t}\chiz\bar{\textrm{t}}\chiz$
    & \ref{fig:t2tt}
    & $-$ & $-$ \\ 
    $D$ &
    $\textrm{pp}\,\rightarrow\,\sBot\sBot\,\rightarrow\,\textrm{b}\chiz\bar{\textrm{b}}\chiz$
    & \ref{fig:t2bb} & $\approx$500 & $\approx$175 \\
    $E$ & $\textrm{pp}\,\rightarrow\,\sGlu\sGlu\,\rightarrow\,\textrm{t}\bar{\textrm{t}}\chiz\textrm{t}\bar{\textrm{t}}\chiz$ & \ref{fig:t1tttt}  & $\approx$850 & $\approx$250 \\
    $F$ &
    $\textrm{pp}\,\rightarrow\,\sGlu\sGlu\,\rightarrow\,\textrm{b}\bar{\textrm{b}}\chiz\textrm{b}\bar{\textrm{b}}\chiz$
    & \ref{fig:t1bbbb} & $\approx$1025 & $\approx$550 \\
    \hline
  \end{tabular}
\end{table}

\begin{figure}[htbp]
  \begin{center}
    \subfigure[\label{fig:t1}$\sGlu\sGlu\,\rightarrow\,\textrm{q}\bar{\textrm{q}}\chiz \textrm{q}\bar{\textrm{q}}\chiz$ (Model $A$)]{
      \includegraphics[width=0.45\textwidth]{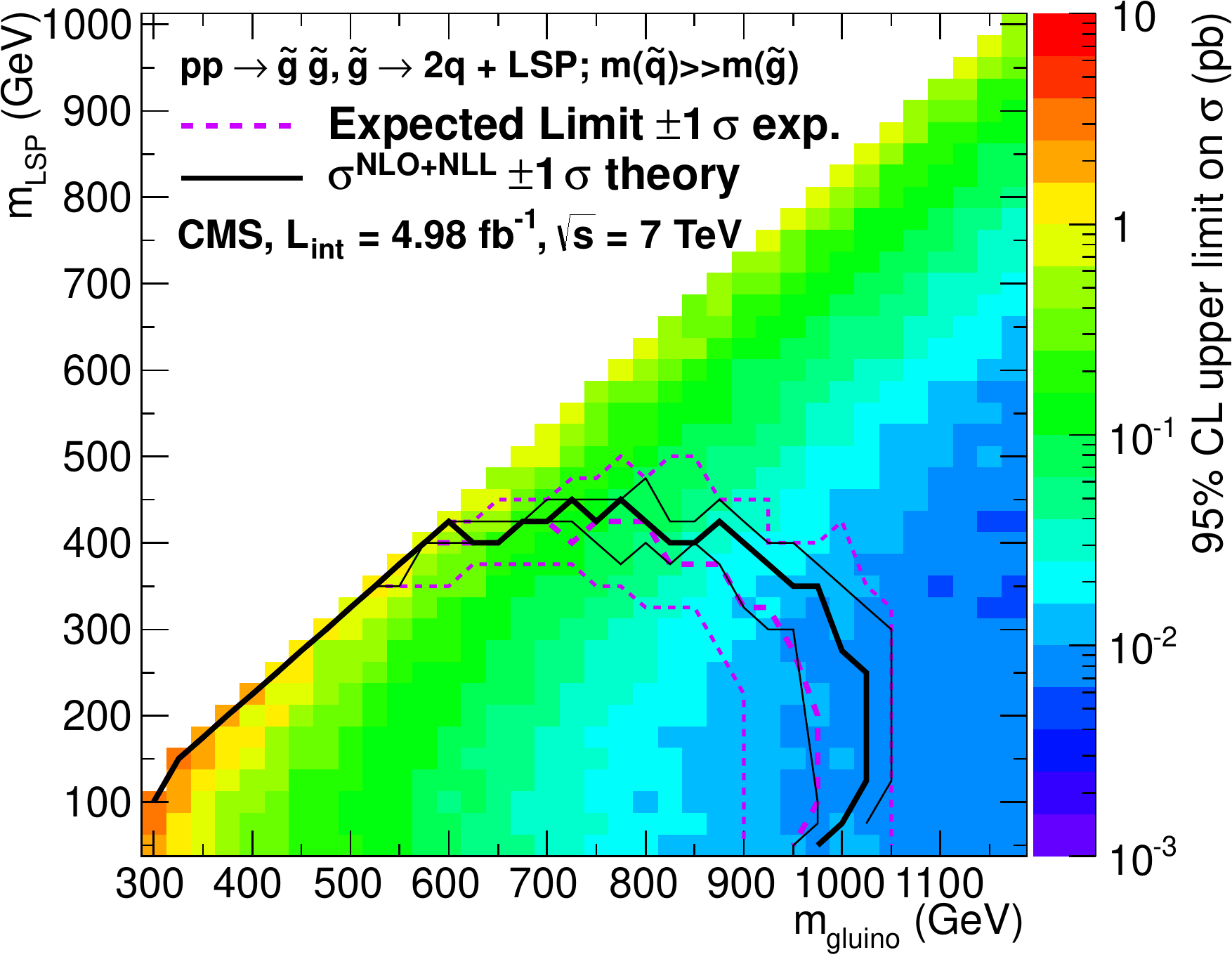}
    } \quad
    \subfigure[\label{fig:t2}$\sQua\sQua\,\rightarrow\,\textrm{q}\chiz \bar{\textrm{q}}\chiz$ (Model $B$)]{
      \includegraphics[width=0.45\textwidth]{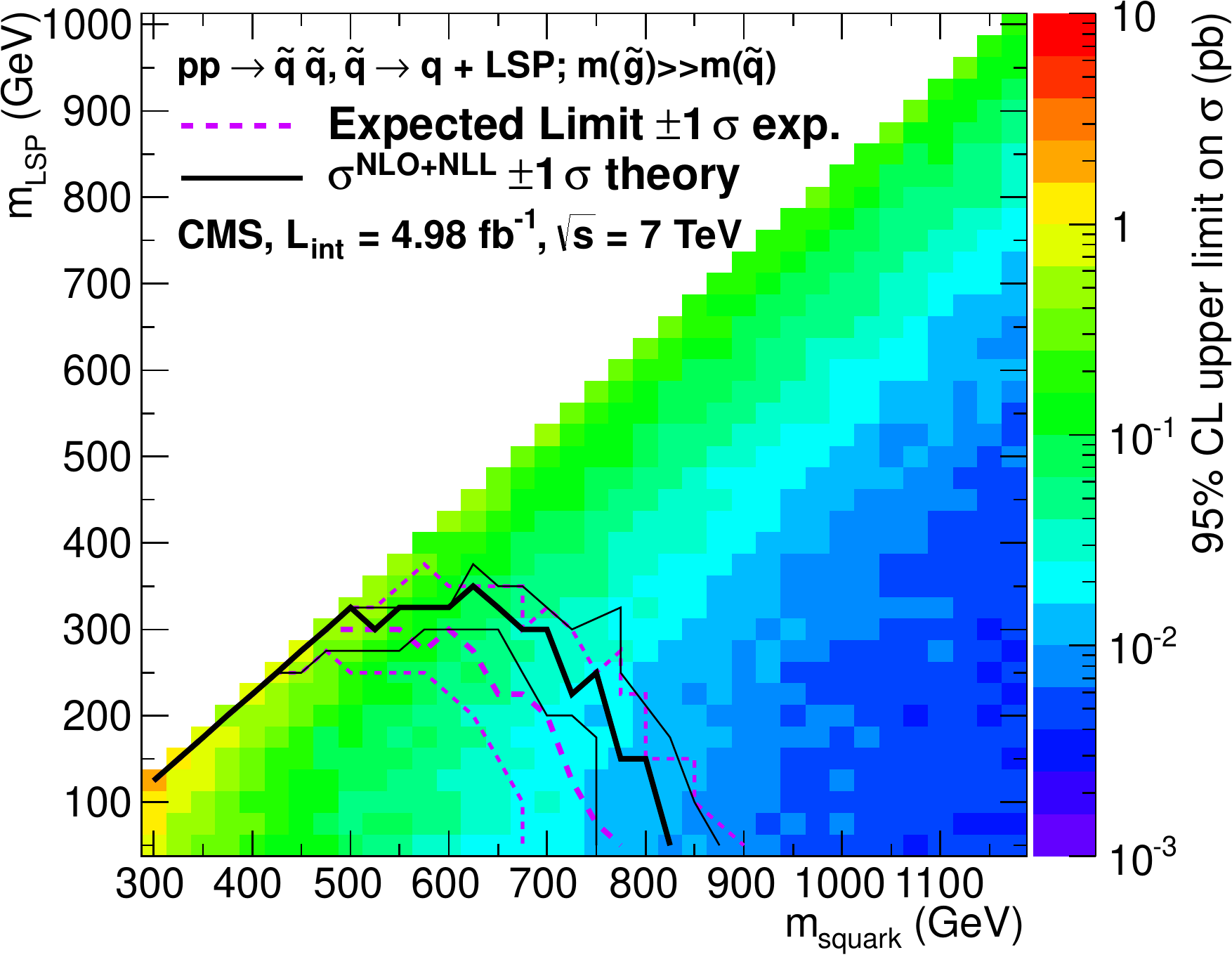}
    } \\
    \subfigure[\label{fig:t2tt}$\sTop\sTop\,\rightarrow\,\textrm{t}\chiz \bar{\textrm{t}}\chiz$ (Model $C$)]{
      \includegraphics[width=0.45\textwidth]{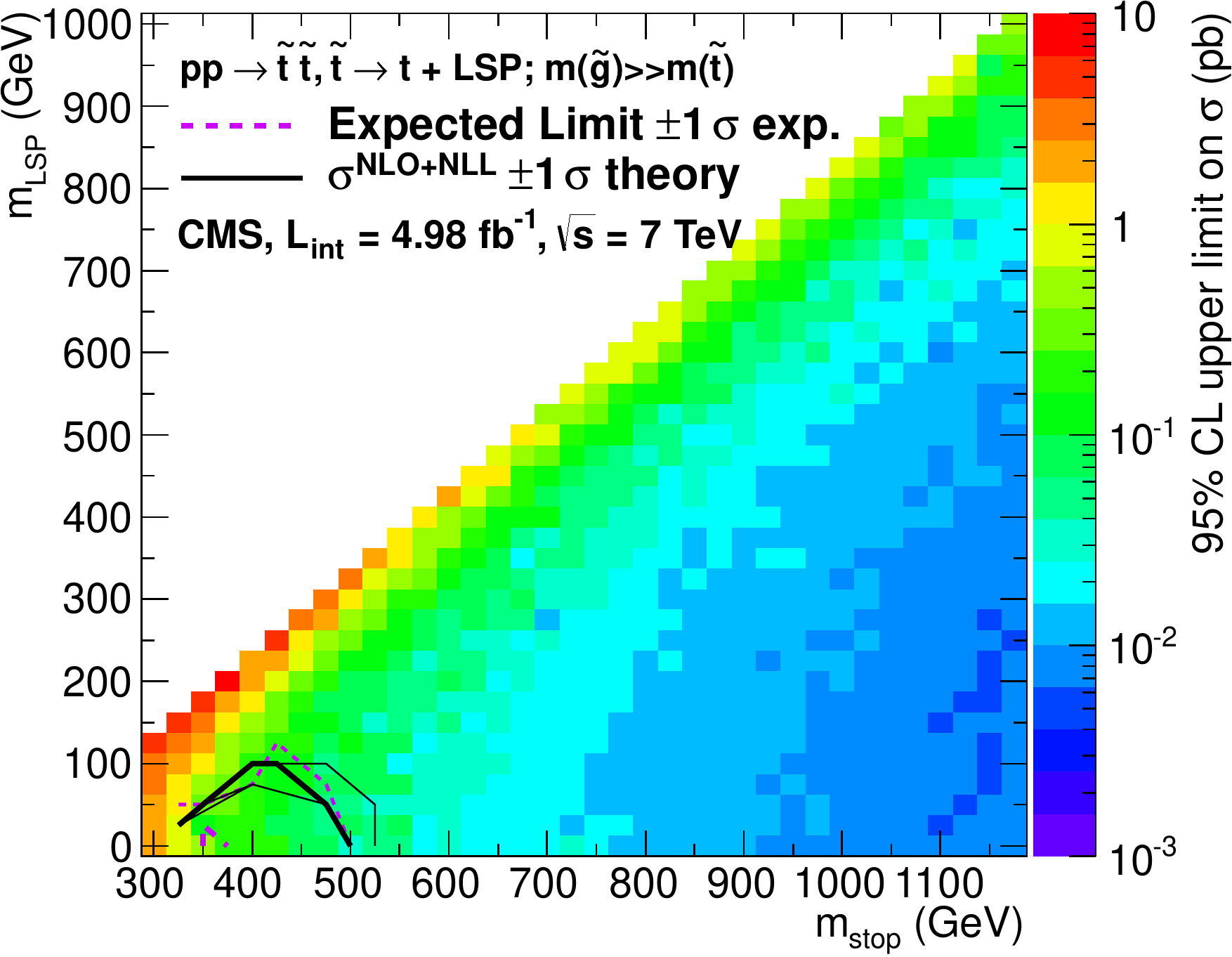}
    } \quad
    \subfigure[\label{fig:t2bb}$\sBot\sBot\,\rightarrow\,\textrm{b}\chiz \bar{\textrm{b}}\chiz$ (Model $D$)]{
      \includegraphics[width=0.45\textwidth]{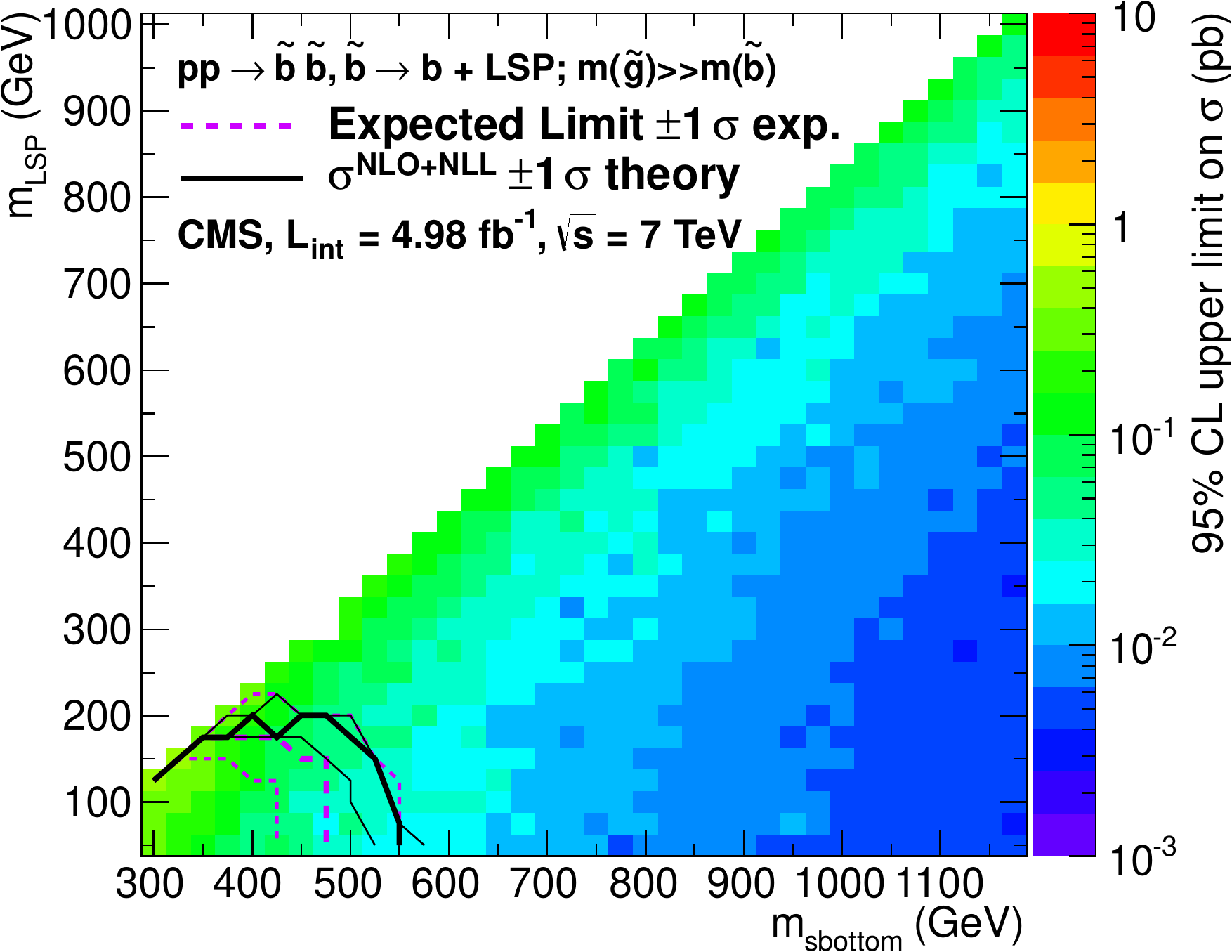}
    } \\
    \subfigure[\label{fig:t1tttt}$\sGlu\sGlu\,\rightarrow\,\textrm{t}\bar{\textrm{t}}\chiz \textrm{t}\bar{\textrm{t}}\chiz$ (Model $E$)]{
      \includegraphics[width=0.45\textwidth]{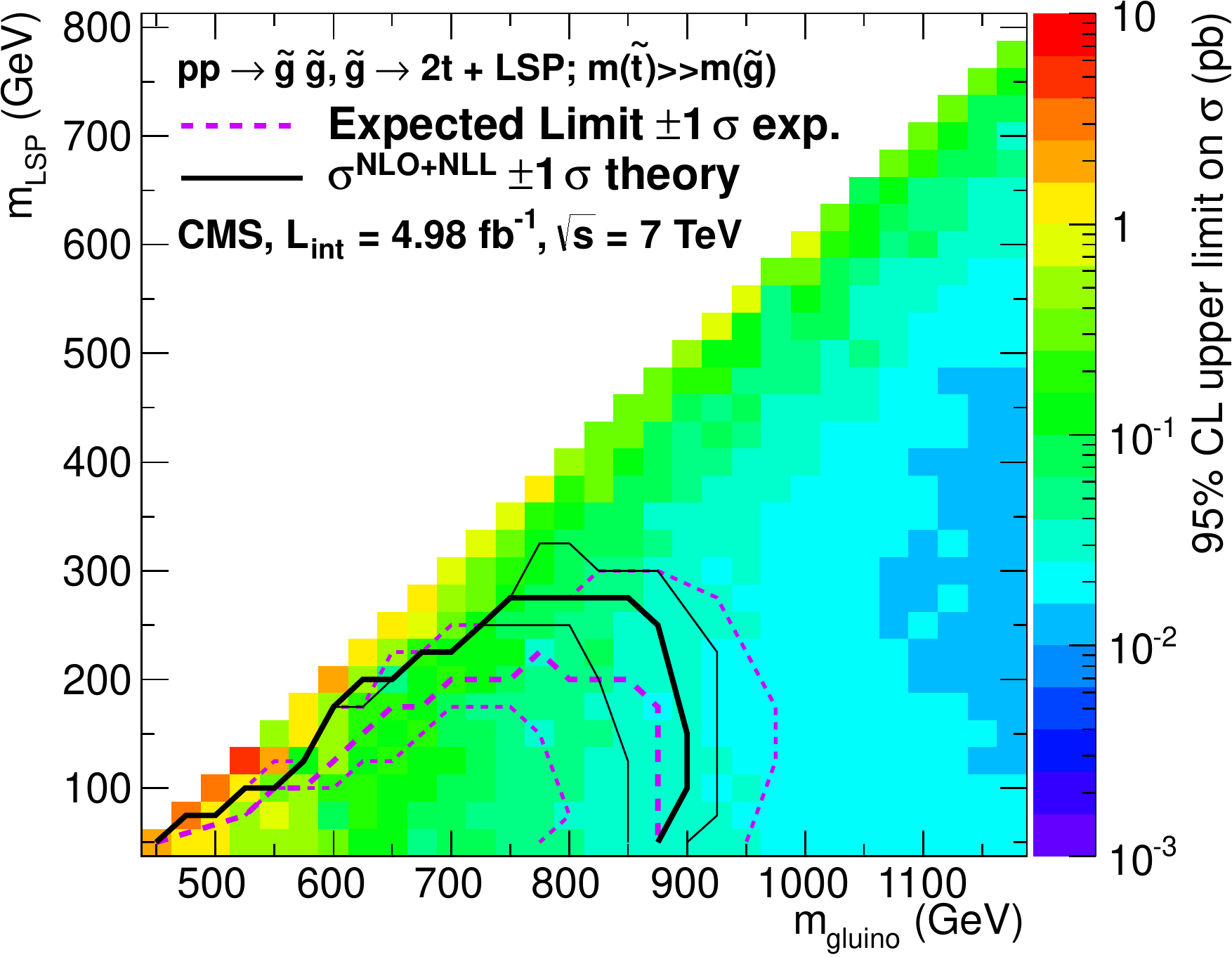}
    } \quad
    \subfigure[\label{fig:t1bbbb}$\sGlu\sGlu\,\rightarrow\,\textrm{b}\bar{\textrm{b}}\chiz \textrm{b}\bar{\textrm{b}}\chiz$ (Model $F$)]{
      \includegraphics[width=0.45\textwidth]{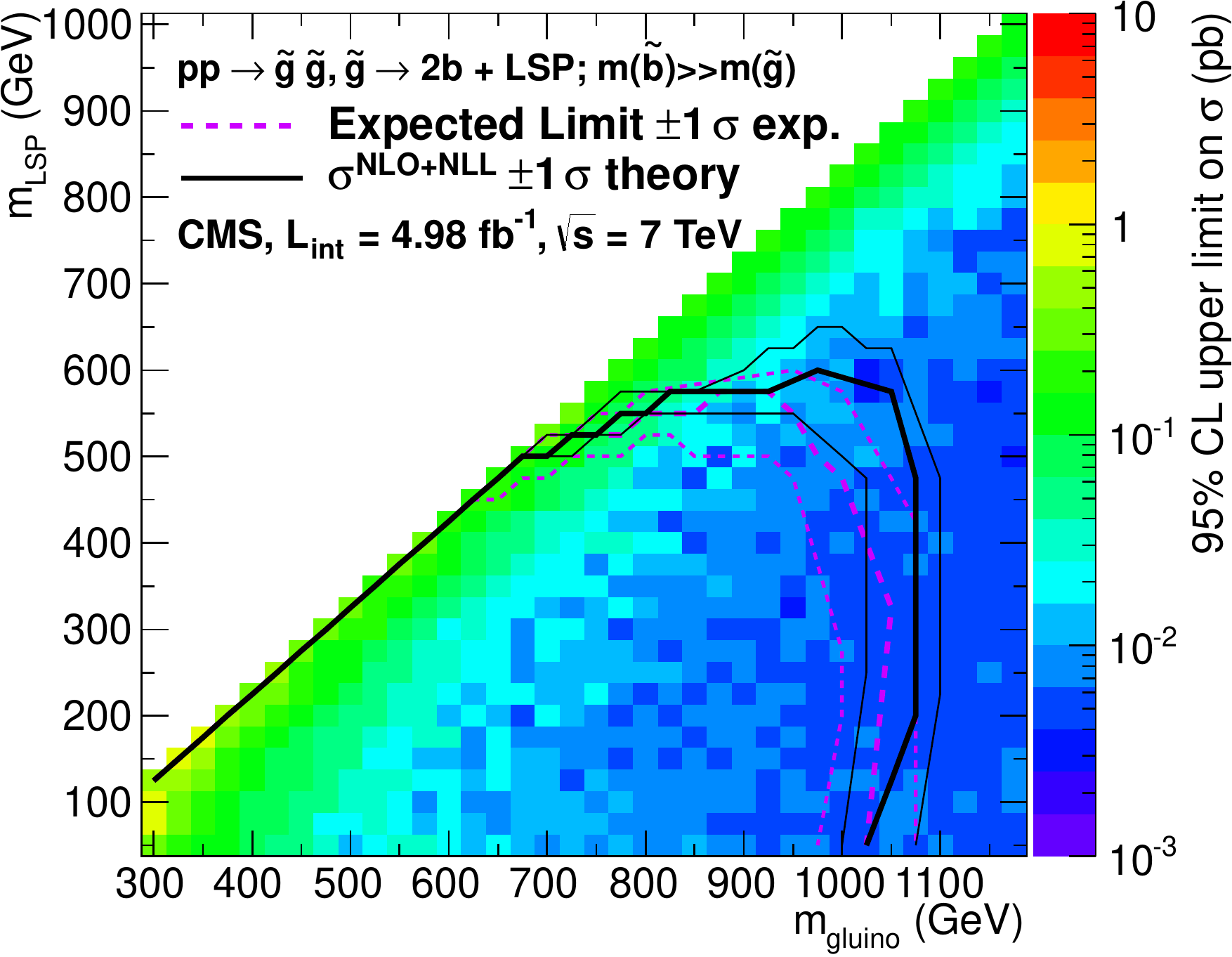}
    } \\
    \caption{\label{fig:limits-sms} Upper limit on cross section at
      95\% CL as a function of $m_{\sq}$ or $m_{\gl}$ and $m_{\rm
        LSP}$ for various simplified models. The solid thick black
      line indicates the observed exclusion region assuming NLO+NLL
      SUSY production cross section. The thin black lines represent
      the observed excluded region when varying the cross section by
      its theoretical uncertainty. The dashed purple lines indicate
      the median (thick line) $\pm 1 \sigma$ (thin lines) expected
      exclusion regions. The mass ranges considered for
      models $C$ and $E$ differ from the other models.}
  \end{center}
\end{figure}

Figure~\ref{fig:limits-sms} shows the upper limit on the cross section
at 95\% CL as a function of $m_{\sq}$ or $m_{\gl}$ and $m_\mathrm{LSP}$
for various simplified models. The point-to-point fluctuations are due
to the finite number of pseudo-experiments used to determine the
observed upper limit. The solid thick black line indicates the
observed exclusion region assuming NLO+NLL 
SUSY cross section for squark pair production in the
limit of very massive gluinos (or vice versa). The thin black lines
represent the observed excluded region when varying the cross section
by its theoretical uncertainty. The dashed purple lines indicate the
median (thick line) ${\pm}1\,\sigma$ (thin lines) expected exclusion
regions.

The most stringent mass limits on the pair-produced sparticles are
obtained at low LSP masses, while the limits typically weaken for
compressed spectra, \ie, points close to the diagonal. In particular,
for all of the considered simplified models, there is an LSP mass
beyond which no limit can be set. This is illustrated in
Figure~\ref{fig:t1}, where the most stringent limit on the gluino mass
is obtained at around $950\GeV$ for low LSP masses, while this limit
weakens to below $900\GeV$ when the LSP mass reaches $350\GeV$. For
LSP masses above $400\GeV$, no gluino masses can be
excluded. Table~\ref{tab:sms} summarises these two extreme cases for
models $A$ to $F$. The estimates on the mass limits are determined conservatively
from the observed exclusion based on the theoretical production cross
section minus $1\sigma$ uncertainty.

\begin{figure}[thb]
  \begin{center}
    \includegraphics[width=0.6\textwidth]{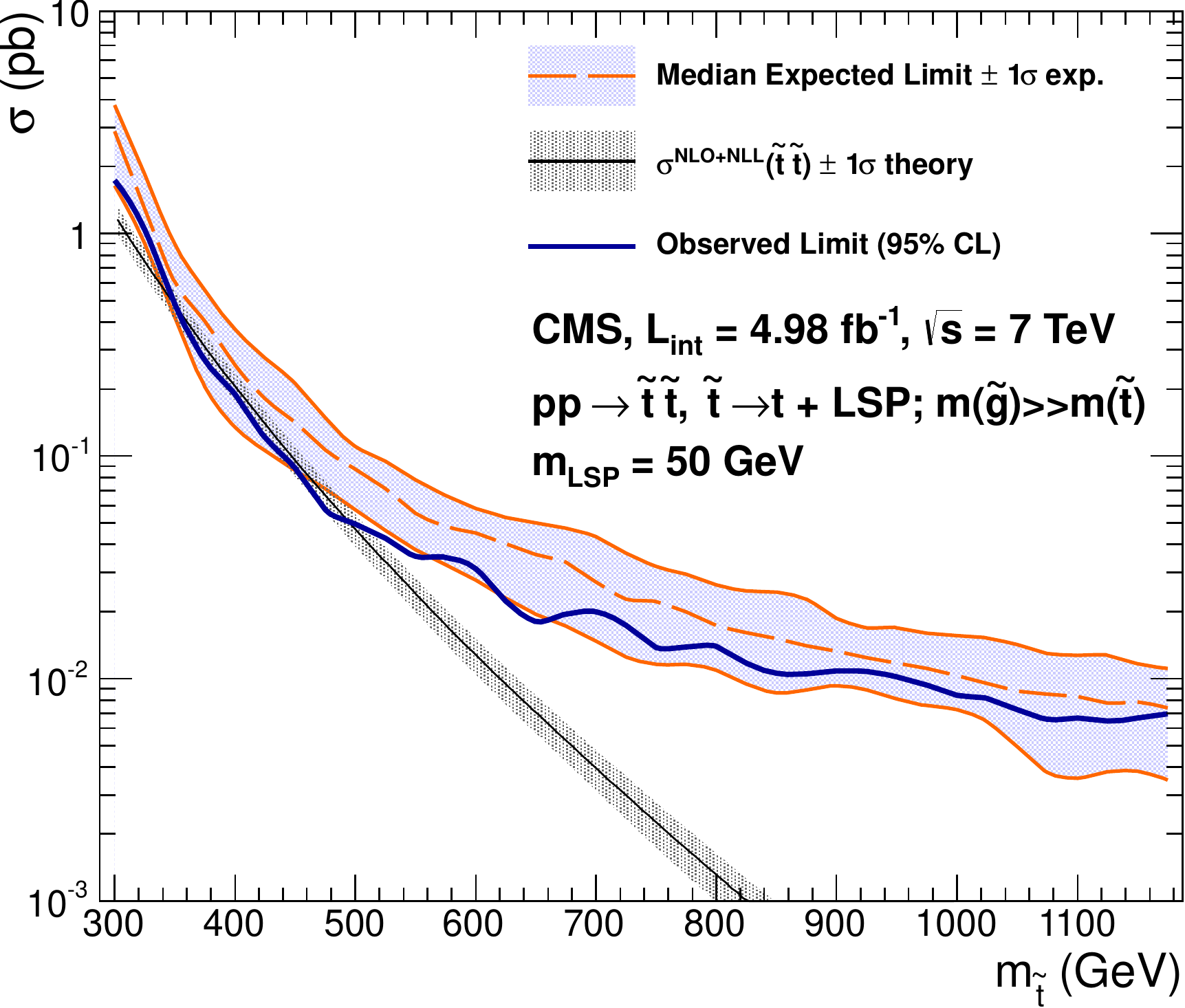}
    \caption{\label{fig:t2tt-mlsp50} Excluded cross section versus top
      squark mass for a model in which pair-produced top squarks decay
      to two top quarks and two neutralinos of mass $m_{\rm LSP} =
      50\gev$. The solid blue line indicates the observed cross
      section upper limit (95\% CL) as a function of the top squark
      mass, $m_{\sTop}$. The dashed orange line and blue band indicate
      the median expected excluded cross section with experimental
      uncertainties. The solid black line and grey band indicate the
      NLO+NLL SUSY top squark pair-production cross section and
      theoretical uncertainties.}
  \end{center}
\end{figure}

No exclusion of direct top squark pair production (model $C$) assuming
the NLO+NLL production cross section is expected with the analysed
dataset and for LSP masses greater than
50\gev. Figure~\ref{fig:t2tt-mlsp50} shows the observed upper limit at
95\% CL on the cross section as a function of the top squark mass
($m_{\sTop}$) only, for a fixed LSP mass of $m_\mathrm{LSP} =
50\gev$. Within the mass range $350 < m_{\sTop} < 475\gev$, the
observed upper limit fluctuates about the theoretical production cross
section minus 1$\sigma$ uncertainty. This mass range is fully excluded
when considering the nominal production cross section.

\section{Summary\label{sec:summary}}

A search for supersymmetry using the CMS detector is reported, based on a data sample of pp
collisions collected at $\sqrt{s} = 7\TeV$, corresponding to an
integrated luminosity of $4.98\pm0.11\fbinv$. Final states with
two or more jets and significant \met, as expected from high-mass
squark and gluino production and decays, have been analysed.  An
exclusive search has been performed in a binned signal region defined
by the scalar sum of the transverse energy of jets, \scalht, and the
number of jets identified to originate from a bottom quark. The sum of
standard model backgrounds per bin has been estimated from a
simultaneous binned likelihood fit to hadronic, $\mu$ + jets, $\mu\mu$
+ jets, and $\gamma$ + jets samples. The observed yields are found to
be in agreement with the expected contributions from standard model
processes. Limits in the CMSSM $(m_0 , m_{1/2})$ plane for $\tan \beta
= 10$, $A_0 = 0\GeV$, and $\mu > 0$ have been derived. For this choice
of parameter values, gluino masses below $700\GeV$ are excluded at
95\% CL. The exclusion increases to $1250\GeV$ for squarks and gluinos
of comparable mass.  Furthermore, exclusion limits are also set in
simplified models, with a special emphasis on third generation squarks
and compressed spectra scenarios. In the considered models with gluino
pair production and for small LSP masses, typical exclusion limits of
the gluino mass are around 1\TeV. For simplified models with squark
pair production, first or second generation squarks are excluded up to
around 750\GeV and bottom squarks are excluded up to around 500\GeV,
again for small LSP masses. No exclusion is expected for direct pair
production of top squarks that each decay to a top quark and a
neutralino of mass $m_{\textrm{LSP}} > 50\gev$. However, within the
mass range $350 < m_{\sTop} < 475\gev$ and for $m_{\textrm{LSP}} =
50\gev$, the observed upper limit fluctuates about the theoretical
production cross section minus 1$\,\sigma$ uncertainty. Thus, for
the simplified models under consideration, the most constraining limits
on the LSP and third-generation squark masses indicate that a large
range of SUSY parameter space is yet to be probed by the LHC.

\section*{Acknowledgements}

\hyphenation{Bundes-ministerium Forschungs-gemeinschaft Forschungs-zentren} We congratulate our colleagues in the CERN accelerator departments for the excellent performance of the LHC and thank the technical and administrative staffs at CERN and at other CMS institutes for their contributions to the success of the CMS effort. In addition, we gratefully acknowledge the computing centres and personnel of the Worldwide LHC Computing Grid for delivering so effectively the computing infrastructure essential to our analyses. Finally, we acknowledge the enduring support for the construction and operation of the LHC and the CMS detector provided by the following funding agencies: the Austrian Federal Ministry of Science and Research; the Belgian Fonds de la Recherche Scientifique, and Fonds voor Wetenschappelijk Onderzoek; the Brazilian Funding Agencies (CNPq, CAPES, FAPERJ, and FAPESP); the Bulgarian Ministry of Education, Youth and Science; CERN; the Chinese Academy of Sciences, Ministry of Science and Technology, and National Natural Science Foundation of China; the Colombian Funding Agency (COLCIENCIAS); the Croatian Ministry of Science, Education and Sport; the Research Promotion Foundation, Cyprus; the Ministry of Education and Research, Recurrent financing contract SF0690030s09 and European Regional Development Fund, Estonia; the Academy of Finland, Finnish Ministry of Education and Culture, and Helsinki Institute of Physics; the Institut National de Physique Nucl\'eaire et de Physique des Particules~/~CNRS, and Commissariat \`a l'\'Energie Atomique et aux \'Energies Alternatives~/~CEA, France; the Bundesministerium f\"ur Bildung und Forschung, Deutsche Forschungsgemeinschaft, and Helmholtz-Gemeinschaft Deutscher Forschungszentren, Germany; the General Secretariat for Research and Technology, Greece; the National Scientific Research Foundation, and National Office for Research and Technology, Hungary; the Department of Atomic Energy and the Department of Science and Technology, India; the Institute for Studies in Theoretical Physics and Mathematics, Iran; the Science Foundation, Ireland; the Istituto Nazionale di Fisica Nucleare, Italy; the Korean Ministry of Education, Science and Technology and the World Class University program of NRF, Korea; the Lithuanian Academy of Sciences; the Mexican Funding Agencies (CINVESTAV, CONACYT, SEP, and UASLP-FAI); the Ministry of Science and Innovation, New Zealand; the Pakistan Atomic Energy Commission; the Ministry of Science and Higher Education and the National Science Centre, Poland; the Funda\c{c}\~ao para a Ci\^encia e a Tecnologia, Portugal; JINR (Armenia, Belarus, Georgia, Ukraine, Uzbekistan); the Ministry of Education and Science of the Russian Federation, the Federal Agency of Atomic Energy of the Russian Federation, Russian Academy of Sciences, and the Russian Foundation for Basic Research; the Ministry of Science and Technological Development of Serbia; the Secretar\'{\i}a de Estado de Investigaci\'on, Desarrollo e Innovaci\'on and Programa Consolider-Ingenio 2010, Spain; the Swiss Funding Agencies (ETH Board, ETH Zurich, PSI, SNF, UniZH, Canton Zurich, and SER); the National Science Council, Taipei; the Thailand Center of Excellence in Physics, the Institute for the Promotion of Teaching Science and Technology and National Electronics and Computer Technology Center; the Scientific and Technical Research Council of Turkey, and Turkish Atomic Energy Authority; the Science and Technology Facilities Council, UK; the US Department of Energy, and the US National Science Foundation.
Individuals have received support from the Marie-Curie programme and the European Research Council (European Union); the Leventis Foundation; the A. P. Sloan Foundation; the Alexander von Humboldt Foundation; the Belgian Federal Science Policy Office; the Fonds pour la Formation \`a la Recherche dans l'Industrie et dans l'Agriculture (FRIA-Belgium); the Agentschap voor Innovatie door Wetenschap en Technologie (IWT-Belgium); the Ministry of Education, Youth and Sports (MEYS) of Czech Republic; the Council of Science and Industrial Research, India; the Compagnia di San Paolo (Torino); and the HOMING PLUS programme of Foundation for Polish Science, cofinanced from European Union, Regional Development Fund.

\bibliography{auto_generated}

\cleardoublepage \appendix\section{The CMS Collaboration \label{app:collab}}\begin{sloppypar}\hyphenpenalty=5000\widowpenalty=500\clubpenalty=5000\textbf{Yerevan Physics Institute,  Yerevan,  Armenia}\\*[0pt]
S.~Chatrchyan, V.~Khachatryan, A.M.~Sirunyan, A.~Tumasyan
\vskip\cmsinstskip
\textbf{Institut f\"{u}r Hochenergiephysik der OeAW,  Wien,  Austria}\\*[0pt]
W.~Adam, E.~Aguilo, T.~Bergauer, M.~Dragicevic, J.~Er\"{o}, C.~Fabjan\cmsAuthorMark{1}, M.~Friedl, R.~Fr\"{u}hwirth\cmsAuthorMark{1}, V.M.~Ghete, J.~Hammer, N.~H\"{o}rmann, J.~Hrubec, M.~Jeitler\cmsAuthorMark{1}, W.~Kiesenhofer, V.~Kn\"{u}nz, M.~Krammer\cmsAuthorMark{1}, I.~Kr\"{a}tschmer, D.~Liko, I.~Mikulec, M.~Pernicka$^{\textrm{\dag}}$, B.~Rahbaran, C.~Rohringer, H.~Rohringer, R.~Sch\"{o}fbeck, J.~Strauss, A.~Taurok, W.~Waltenberger, G.~Walzel, E.~Widl, C.-E.~Wulz\cmsAuthorMark{1}
% ...
\vskip\cmsinstskip
\dag:~Deceased\\
1:~~Also at Vienna University of Technology, Vienna, Austria\\
% ...

\end{sloppypar}
\end{document}